\input harvmac.tex
\input epsf.tex


\def\ha{{\textstyle{1\over2}}}
\def\bar{\overline}

\def\darr#1{\raise1.8ex\hbox{$\leftrightarrow$}\mkern-19.8mu #1}
\def\roughly#1{\ \lower1.5ex\hbox{$\sim$}\mkern-22.8mu #1\,}

\hyphenation{di-men-sion di-men-sion-al di-men-sion-al-ly}

\parindent=0pt
\parskip=5pt

\def\Tr{{\rm Tr}}

\def\zb{{\bar z}}

\def\IR{{\hbox{{\rm I}\kern-.2em\hbox{\rm R}}}}
\def\IB{{\hbox{{\rm I}\kern-.2em\hbox{\rm B}}}}
\def\IN{{\hbox{{\rm I}\kern-.2em\hbox{\rm N}}}}
\def\IC{{\hbox{{\rm I}\kern-.6em\hbox{\bf C}}}}
\def\IZ{{\hbox{{\rm Z}\kern-.4em\hbox{\rm Z}}}}

\def\ap{\alpha^\prime}
\noblackbox

\writetoc

\Title{\vbox{\baselineskip12pt
\hbox{NSF-ITP-96-003}
\hbox{hep-th/9602052}}}
{Notes on D-Branes}

\vskip0.5cm
\centerline{{\bf Joseph Polchinski,
Shyamoli Chaudhuri, Clifford V. Johnson}}
\vskip1.5cm
\baselineskip=11pt
\centerline{\sl Institute For Theoretical Physics}
\centerline{\sl  University of California}
\centerline{\sl Santa Barbara, CA 93106-4030 USA}
\centerline{email: joep,sc,cvj@itp.ucsb.edu}
\vfill
\centerline{\bf Abstract}
\vskip0.5cm
\vbox{\narrower\baselineskip=12pt\noindent
This is a series of remedial lectures on open and unoriented strings
for the heterotic string generation.  The particular focus is on the
interesting features that arise under $T$-duality---D-branes and
orientifolds.  The final lecture discusses the application to string
duality. There will be no puns. Lectures presented by J. P. at the
ITP from Nov.~16 to Dec.~5, 1995.  References updated through Jan.~25,
1996.}

\Date{January 1996}



\centerline{\bf Contents}
\vskip1cm
\noindent {1.} {Lecture I: {\fam \slfam \tensl Open and Unoriented Bosonic
Strings} }  \vfill \par
\noindent \quad{1.1.} {Open Strings} \vfill\par
\noindent \quad{1.2.} {Chan-Paton Factors} \vfill \par
\noindent \quad{1.3.} {Unoriented Strings} \vfill \par
\noindent {2.} {Lecture II: {\fam \slfam \tensl $T$-Duality}} \vfill \par
\noindent \quad{2.1.} {Self Duality of Closed Strings} \vfill \par
\noindent \quad{2.2.} {Open Strings and Dirichlet-Branes} \vfill \par
\noindent \quad{2.3.} {Chan-Paton Factors and Multiple D-Branes} \vfill \par
\noindent \quad{2.4.} {D-Brane Dynamics} \vfill \par
\noindent \quad{2.5.} {D-Brane Tension} \vfill \par
\noindent \quad{2.6.} {Unoriented Strings and Orientifolds.} \vfill \par
\noindent {3.} {Lecture III: {\fam \slfam \tensl Superstrings and $T$-Duality}}
\vfill \par
\noindent \quad{3.1.} {Open Superstrings} \vfill \par
\noindent \quad{3.2.} {Closed Superstrings} \vfill \par
\noindent \quad{3.3.} {$T$-Duality of Type II Superstrings} \vfill \par
\noindent \quad{3.4.} {$T$-Duality of Type I Superstrings} \vfill \par
\noindent {4.} {Lecture IV: {\fam \slfam \tensl D-Branes Galore}} \vfill\par
\noindent \quad{4.1.} {Discussion} \vfill \par
\noindent \quad{4.2.} {Multiple Branes and Broken Supersymmetries} \vfill \par
\noindent \quad{4.3.} {D-strings} \vfill \par
\noindent \quad{4.4.} {Five-Branes} \vfill \par
\noindent \quad{4.5.} {A Brief Survey} \vfill \par
\noindent \quad{4.6.} {Conclusion} \vfill \par

\vfill\eject

\baselineskip13pt

\nref\dbranes{J.~Dai, R.~G.~Leigh and J.~Polchinski, Mod.~Phys.~Lett.
A4 (1989) 2073;\hfil\break R.~G.~Leigh, Mod.~Phys.~Lett. A4 (1989) 2767.}
\nref\joeone{ J.~Polchinski, Phys.~Rev.~D50 (1994) 6041.}
\nref\joetwo{ J.~Polchinski, Phys.~Rev. Lett.~{\bf 75} (1995) 4724.}

String theory today is a bit like particle physics in
the good old days: we have a great deal of `data' coming in, and are
looking for the theory that explains it.  Of course the data is not
experimental but theoretical.  For the most part, it consists of
evidence that all the different string theories and string backgrounds
that have been found are different states in a single theory.
For many years students
were told that it was sufficient to study closed oriented strings,
the heterotic string in particular, but now various strongly coupled
limits of the heterotic string theory are weakly coupled open or
unoriented theories.  Moreover, certain solitonic states, required by
string duality, turn out to have a simple interpretation in terms of open
strings with exotic boundary conditions.

These lectures are thus intended as a remedial course in open and unoriented
strings and the exotic things that happen to them under $T$-duality, in
particular the appearance of orientifolds and D-branes. We will start with
the bosonic string, as many of the interesting features already appear
there,  but
some of the essential structure will arise only in the supersymmetric case.
The presentation is largely an expanded version of
refs.~\refs{\dbranes,\joeone,\joetwo}, with a few of the more recent
developments.  In addition to references at appropriate points in the text,
we will include at the end a short survey of the literature on this subject,
both pre- and post-string duality. A general familiarity with string theory
is assumed.

\newsec{Lecture I: {\sl Open and Unoriented Bosonic Strings} }

\subsec{Open Strings}
To parameterize the open string's world sheet, let the `spatial' coordinate
run $0 \leq \sigma^1 \leq \pi$ as in the figure:

\vskip1.5cm
\hskip2.5cm\epsfxsize=2.5in\epsfbox{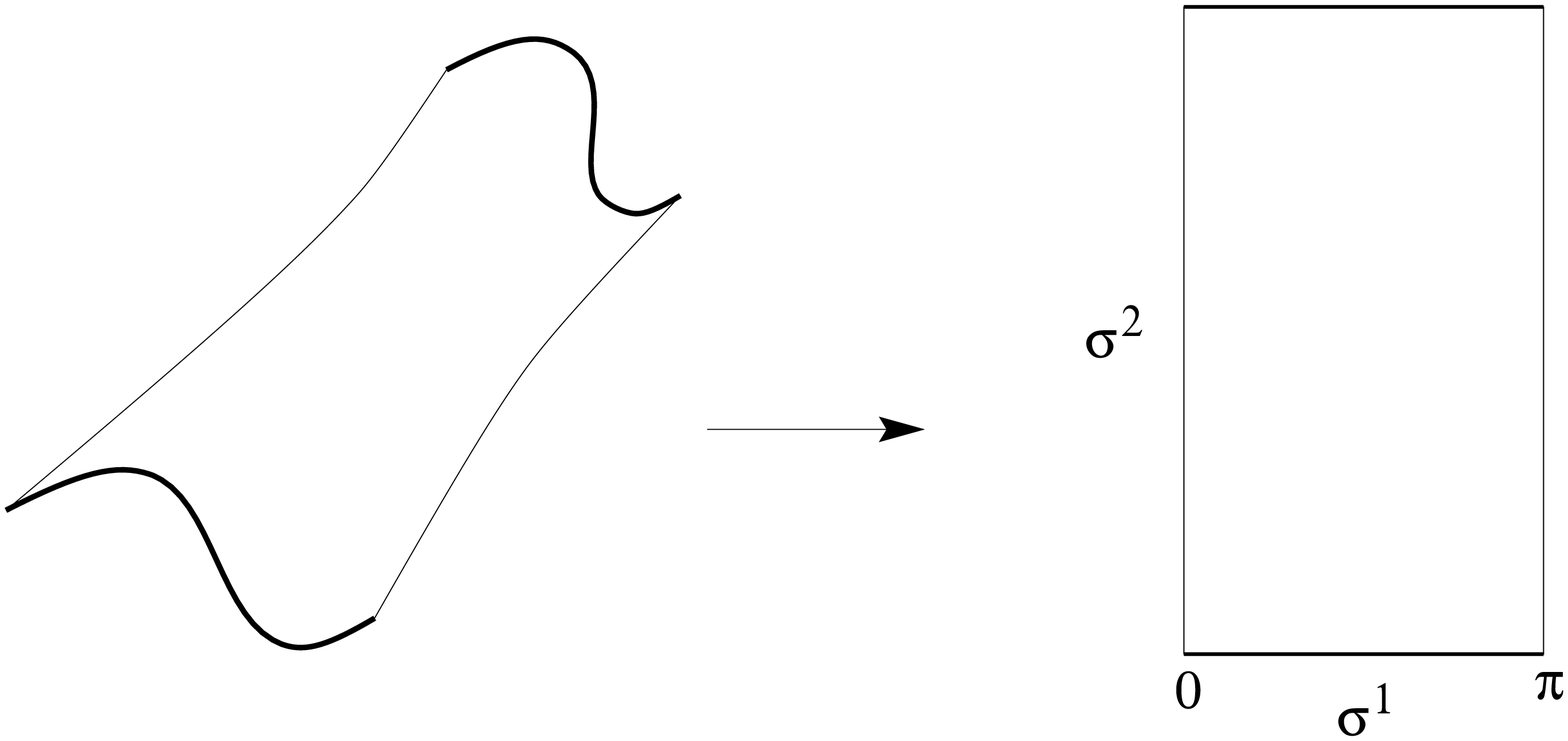}
\vskip0.4cm

In conformal gauge, we have the action:
\eqn\action{{1\over4\pi\ap}\int_{\cal M}d^2\sigma
\left(\partial_aX^\mu\partial_aX_\mu\right)}
Under a variation, after integrating by parts,
\eqn\variation{\delta S=-{1\over2\pi\ap}\int_{\cal M}d^2\sigma \left(\delta
X^\mu\partial^2
X_\mu\right)+{1\over2\pi\ap}\int_{\partial M}d\sigma\left(\delta
X^\mu\partial_nX_\mu\right)}
where $\partial_n$ is the derivative normal to the boundary.
The only Poincar\'e invariant boundary condition is the
Neumann condition
$\partial_n X_\mu=0$.  The Dirichlet condition $X^\mu= {\rm constant}$
is also consistent with the equation of motion, and we might study it for
its own sake.  However, we will follow history and begin with the Neumann
condition, finding that Dirichlet condition
is forced upon us later.
{}From the first term in \variation, we have to simply solve Laplace's
equation (or the wave equation, if we have Minkowski signature on the
world-sheet.)

The general solution to Laplace's equation with Neumann boundary conditions
is
\eqn\modexp{X^\mu(z,\zb)=x^\mu-i\ap p^\mu
\ln(z\zb)+i\sqrt{\ap\over2}\sum_{m\neq0}{\alpha_m^\mu\over
m}(z^{-m}+\zb^{-m}),}
where $x^\mu$ and $p^\mu$ are the position and momentum of the center of
mass.  As is conventional in conformal field theory, this has been
written in terms of the coordinate $z=e^{\sigma^2+i\sigma^1}$, so that
time runs radially:

\vskip1.5cm
\hskip2.5cm\epsfxsize=2.5in\epsfbox{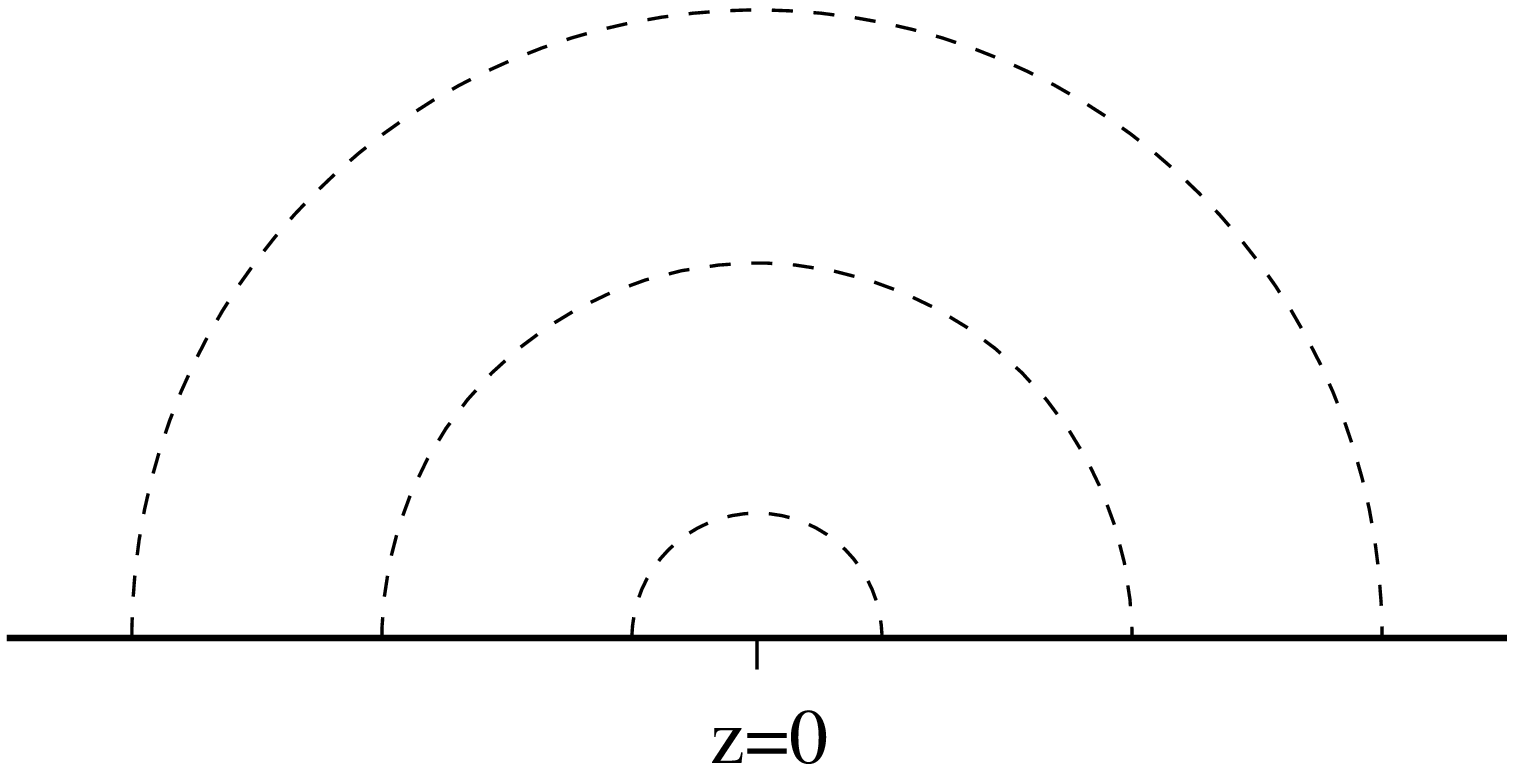}
\vskip0.4cm

After the usual canonical quantization:
\eqn\canon{\eqalign{[x^\mu,p^\nu]&=i\eta^{\mu\nu};\cr
 [\alpha_m^\mu,\alpha_n^\nu]&=m\delta_{m+n}\eta^{\mu\nu},}}
and  we get the mass spectrum
\eqn\masses{M^2=-p^\mu p_\mu={1\over\ap}\left(\sum_{m=1}^\infty
mN_m-1\right).} Here $N_m$ is the number of excited oscillators in the
$m^{th}$ mode, and the $-1$ is the zero point energy of the physical
bosons.\foot{A boson with periodic boundary conditions has zero point energy
$-{1\over24}$, and with antiperiodic boundary conditions it is $1\over48$.
For fermions, there is an extra minus sign. For the bosonic string in 26
dimensions, there are 24 transverse (physical) degrees of freedom.}

So for example we have  the following two particle states:
\eqn\particles{\matrix{&\rm tachyon:&
|{ k}\rangle,& M^2=-{1\over\ap}, & V=\exp({i{ k}\cdot { X}});\cr
&\rm photon:&\alpha^\mu_{-1}|{ k}\rangle,
& M^2 =0,&V^\mu=\partial_t X^\mu
\exp({i{ k}\cdot{ X}}),}} where $V$ is the particle state's vertex
operator. Here $\partial_t$
is the derivative tangent to the string's world sheet boundary.

\subsec{Chan-Paton Factors}
It is consistent with spacetime Poincar\'e invariance and world-sheet
conformal invariance to add non-dynamical degrees of freedom at the ends.
Their Hamiltonian vanishes so these degrees of freedom
have no dynamics---an end of the string prepared in one of these states will
remain in one of these states. So in addition to the Fock space label for the
string,
we could label each end $i$ or $j$ where the labels run from $1$ to
$N$:\ref\chanpat{J. Paton and Chan Hong-Mo, Nucl. Phys. {\bf B10} (1969)
519.}

\vskip0.5cm
\hskip4.5cm\epsfxsize=1.5in\epsfbox{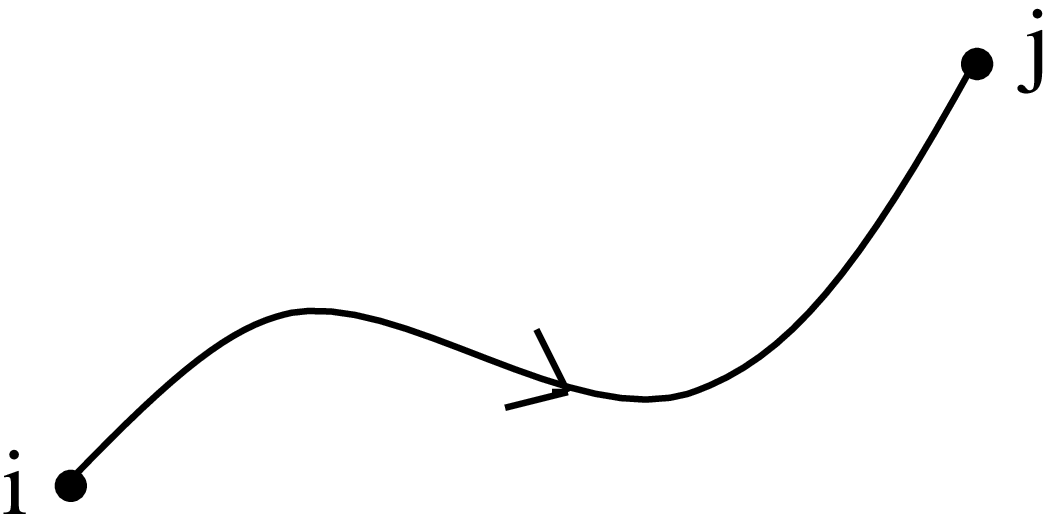}
\vskip0.4cm

The $n\times n$ matrix $\lambda^a_{ij}$ forms a basis into which to
decompose a string wavefunction $|{ k},a\rangle=\sum^N_{i,j}|{
k},ij\rangle\lambda^a_{ij}$. These are the Chan-Paton factors. Each vertex
operator carries such a factor.
String fields satisfy a reality condition (e.g. the graviton must be real),
so for the Chan-Paton factors $\lambda_{ij}=\lambda^*_{ji}$---they are
Hermitian.  Later we shall see there are extra conditions on the
$\lambda$ arising from requiring certain factorization properties of
amplitudes, and one-loop consistency of the theory. The tree
diagram for four oriented strings is:

\vskip1.0cm
\hskip2.0cm\epsfxsize=3.0in\epsfbox{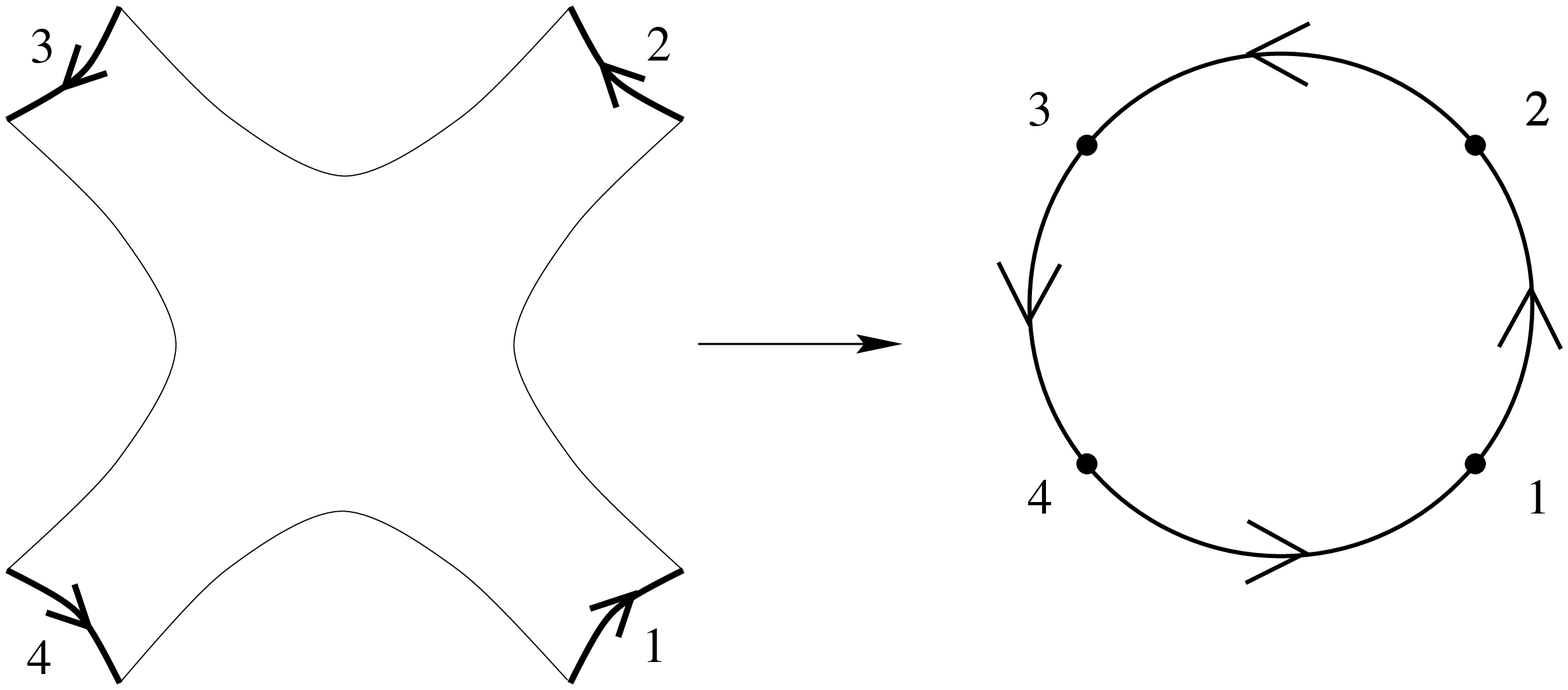}
\vskip0.4cm

Since the Chan-Paton degrees of freedom are non-dynamical, the right end of
string 1 must have the same index as the left end of string 2, and so on,
and so the diagram comes with a factor
\eqn\factor{\lambda^1_{ij}\lambda^2_{jk}\lambda^3_{kl}
\lambda^4_{li}=\Tr(\lambda^1\lambda^2\lambda^3\lambda^4).}
So in general, we multiply by such factors in order to account for  the
Chan-Paton degrees of freedom in amplitudes. The massless vertex operator
$V^{a\mu}= \lambda^a_{ij}\partial_tX^\mu\exp(i{ k}\cdot { X})$
transforms as the adjoint under the $U(N)$ symmetry of the Chan-Paton
degrees of freedom (under which the endpoints transform as an $\bf N$ and
$\bf \bar N$ respectively), so this is a gauge symmetry in spacetime.

The massless bosonic background fields of the open string are the graviton
$G_{\mu\nu}$, dilaton $\phi$, and antisymmetric tensor $B_{\mu\nu}$. The
closed string coupling is related to the expectation value of the dilaton
field
$\phi_0$ and is given by $g=e^{\phi_0}$. In the  low energy limit, the
massless fields satisfy equations of motion which may be obtained by
varying the following action:
\eqn\lowlimit{S=\int
d^{10}X\left\{{1\over2}e^{-2\phi}\left(R+4(\nabla\phi)^2-{1\over12}
H_{\mu\nu\kappa}H^{\mu\nu\kappa}\right)-{c\over 4}e^{-\phi}\Tr
F_{\mu\nu}F^{\mu\nu} +O(\ap)\right\}}
This action arises from tree level in string perturbation theory---the
closed string kinetic terms are accompanied by $g^{-2}$, from the sphere,
and the open string kinetic terms by $g^{-1}$, from the disk.\foot{In string
perturbation theory world-sheets contribute a factor $g^{2h-2+b+c}$, where
$h,b$ and $c$ (which completely characterize the topology of a
two-dimensional manifold) are the number of handles, boundaries and
crosscaps, respectively.} Each further order in $\ap$
brings two extra derivatives and terms  such as $\ap e^{-2\phi}R^2$ and $\ap
e^{-\phi}\Tr[F_\mu^{\phantom{\mu}\nu}
F_\nu^{\phantom{\mu}\lambda}F_\lambda^{\phantom{\mu}\mu}]$ appear.
Some of these terms vanish  in the supersymmetric case.
The normalization of the open string action will be discussed
later.

\subsec{Unoriented Strings}

Let us begin with the open string sector.  World sheet parity acts as the
$z\leftrightarrow-\bar z$, reflecting right-moving modes into
left moving modes. In terms of the mode expansion, $X^\mu(z,\bar z) \to
X^\mu(-\bar z,- z)$ takes
\eqn\have{\eqalign{x^\mu\rightarrow x^\mu\cr
p^\mu\rightarrow p^\mu\cr
\alpha^\mu_m\rightarrow(-1)^m\alpha^\mu_m\ .}}
This is a global symmetry of the open string theory above, but we can also
consider the theory in which it is gauged.  When a discrete symmetry
is gauged, only invariant states are left in the spectrum.\foot{The
familiar example of this is the orbifold construction, in which some
global world-sheet symmetry, usually a discrete symmetry of spacetime, is
gauged.}
The open string tachyon is even and survives the
projection, while the photon does not, as
\eqn\photon{\eqalign{
&\Omega|{ k}\rangle = +|{ k}\rangle;\cr
&\Omega\alpha^\mu_{-1}|{ k}\rangle =  -\alpha^\mu_{-1}|{ k}\rangle.}}
We have made an assumption here about the overall sign of $\Omega$.  This
sign is fixed by the requirement that $\Omega$ be conserved in string
interactions, which is to say that it is a symmetry of the operator
product expansion (OPE).  The assignment~\have\ matches the symmetries
of the vertex operators~\particles: the minus sign is from the
orientation reversal on the tangent derivative $\partial_t$.

World-sheet parity reverses the Chan-Paton factors on the two ends of the
string, but more generally it may have some additional action on each
endpoint,
\eqn\cpomega{\eqalign{\Omega\lambda_{ij}|{ k},ij \rangle\to
\lambda^\prime_{ij} |{ k},ij \rangle\cr
\lambda^\prime=M\lambda^{T}N.
}}
Further it must be that $M = N^{-1}$ in order that this be a symmetry of
general amplitudes such as~\factor.

Acting twice with $\Omega$ squares to the identity on the fields, leaving
only the action on the Chan-Paton degrees of freedom. States are thus
invariant under
\eqn\cpcondition{\lambda\to MM^{-T}\lambda M^TM^{-1}.}
The $\lambda$ must span a complete set of $N \times N$ matrices: if
strings $ik$ and $jl$ are in the spectrum for {\it any} values of $k$ and
$l$, then so is the state $ij$.  First, $jl$ implies $lj$ by CPT, and a
splitting-joining interaction in the middle gives $ik + lj \to ij + lk$.
But now Schur's lemma requires $MM^{-T}$ to be proportional to the
identity,
so $M$ is either symmetric or antisymmetric.
This gives
two cases, up to choice of basis:\ref\sosp
{J. H. Schwarz, in {\it Florence 1982, Proceedings, Lattice Gauge Theory,
Supersymmetry and Grand Unification,} 233; Phys. Rept. {\bf 89} (1982)
223;\hfil\break
N. Marcus and A. Sagnotti, Phys. Lett. {\bf 119B} (1982) 97.}
\item{\bf a.}{ $M=M^T=I_N$. (Here, $I_N$ is the
$N\times N$ unit matrix.) In this case, for the photon
$\lambda_{ij}\alpha^\mu_{-1}|{ k}\rangle$ to be even under $\Omega$ and
therefore survive the projection, we must have $\lambda=-\lambda^T$, which
means that our gauge group is $SO(N)$.}
\item{\bf b.}{$M=-M^T=i\pmatrix{0&I_{N/2}\cr-I_{N/2}&0}$.
In this case, $\lambda=-M\lambda^TM$, which defines the gauge
group\foot{In the notation where $USp(2)\equiv SU(2)$.} $USp(N)$.}

Now consider the closed string sector. For closed strings, we have the
familiar mode expansion
$X^\mu(z,\zb)=X^\mu(z)+X^\mu(\zb)$ with:
\eqn\closedmodes{\eqalign{X^\mu(z)=x^\mu+i\sqrt{\ap\over2}
\left(-\alpha_0^\mu
\ln z+\sum_{m\neq0}{\alpha^\mu_m\over mz^m}\right),\cr
X^\mu(\zb)={\tilde x}^\mu+i\sqrt{\ap\over2}\left(-{\tilde\alpha}_0^\mu
\ln \zb+\sum_{m\neq0}{{\tilde\alpha}^\mu_m\over m\zb^m}\right).}}
The theory is invariant under a  world-sheet parity symmetry
$\sigma^1\to-\sigma^1$.
For a closed string, the action of $\Omega$ is to reverse the right- and
left-moving oscillators:
\eqn\act{\Omega:\quad\alpha^\mu_m\leftrightarrow{\tilde\alpha}^\mu_m.}
For convenience, parity is here taken to be $z \to \bar z$, differing by a
$\sigma^1$-translation from $z \to -\bar z$.  This is a global symmetry,
but again we can gauge it.  We have
${\Omega|{ k}\rangle=|{ k}\rangle}$, and so the tachyon remains in the
spectrum.  However
\eqn\invtwo{\Omega\alpha^\mu_{-1}{\tilde\alpha}^\nu_{-1}|{
k}\rangle= {\tilde\alpha}^\mu_{-1}\alpha^\nu_{-1}|{ k}\rangle,}
so only states symmetric under $\mu\leftrightarrow\nu$ survive from this
multiplet, i.e. the graviton and dilaton. The antisymmetric tensor is
projected out.

When a world-sheet symmetry is gauged,
a string carried around a closed curve on the world-sheet need only come
back to itself up to a gauge transformation.  Gauging world-sheet parity
thus implies the inclusion of unoriented world-sheets.
The oriented one-loop closed string amplitude comes only from the
torus, while insertion of the projector
${1\over2}\Tr(1+\Omega)$ into a closed string one-loop
amplitude will give the amplitude on the torus and Klein bottle
respectively:

\vskip1.0cm
\hskip1.5cm\epsfxsize=3.5in\epsfbox{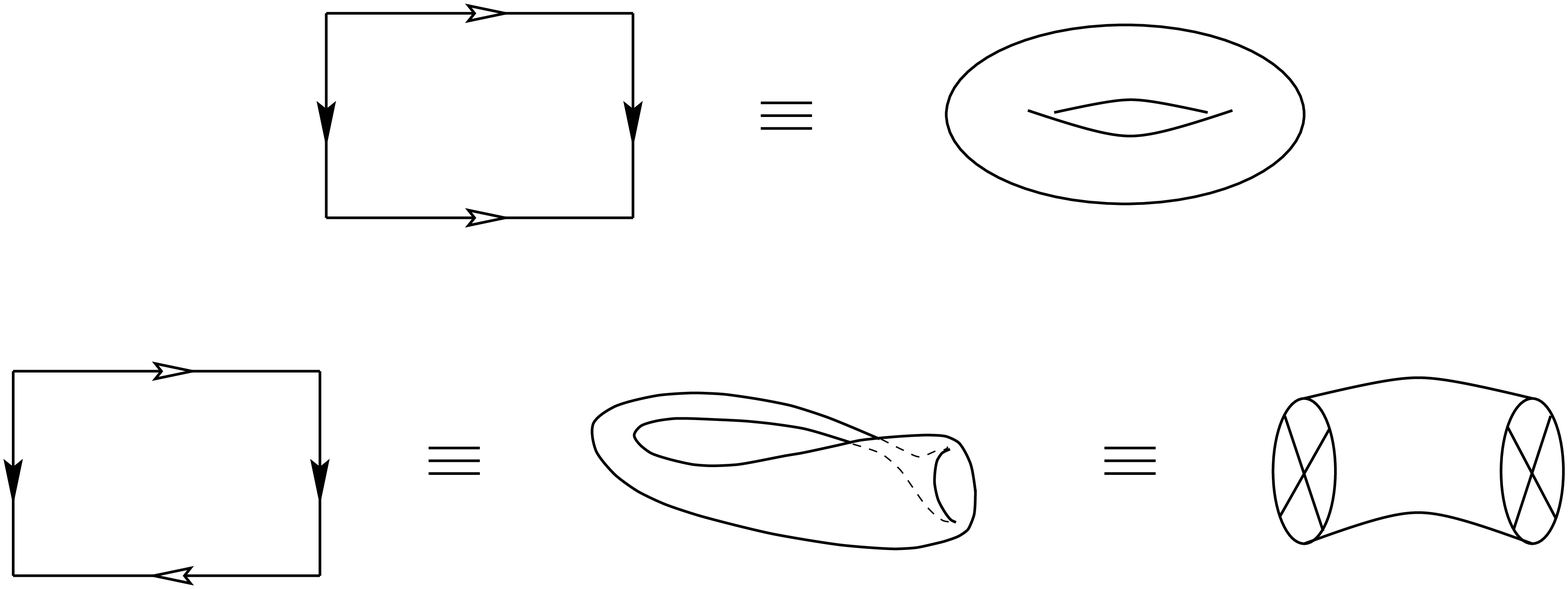}
\vskip0.4cm

Similarly, the unoriented one-loop open string amplitude comes from the
annulus and M\"obius strip.  The lowest order unoriented amplitude is the
projective plane, which is a disk with opposite points identified:

\vskip0.5cm
\hskip4.0cm\epsfxsize=2.0in\epsfbox{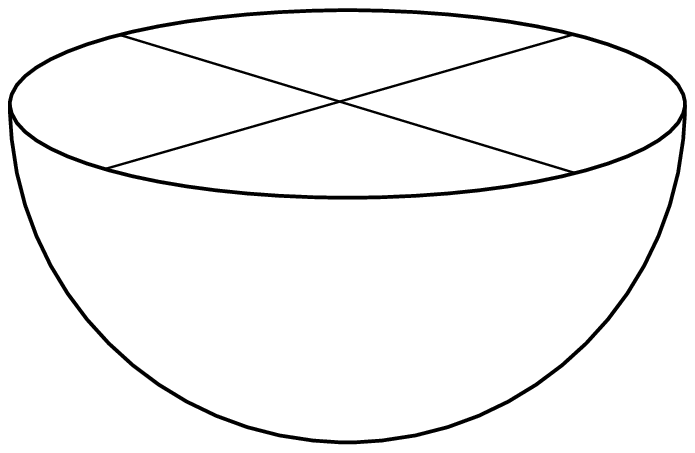}.
\vskip0.4cm

A circular hole with opposite points identified is a crosscap.  The Klein
bottle can be represented as a cylinder with a crosscap at each end,\
as shown in the figure above.
This representation will be useful and will be explained further in
section~2.6.
\nref\ori{A. Sagnotti, in {\it
Non-Perturbative Quantum Field Theory,}
eds. G. Mack et. al. (Pergamon Press, 1988) 521;\hfil\break
G. Pradisi and A. Sagnotti, Phys. Lett. {\bf B216} (1989) 59;\hfil\break
M. Bianchi and A. Sagnotti, Phys. Lett. {\bf 247B}
(1990) 517; Nucl. Phys. {\bf B361} (1991) 519;\hfil\break
P. Horava, Nucl. Phys. {\bf B327} (1989) 461;
Phys. Lett. {\bf B289} (1992) 293; Nucl. Phys. {\bf B418}
(1994) 571; {\it Open Strings from Three Dimensions: Chern-Simons-Witten
Theory an Orbifolds,} Prague preprint PRA-HEP-90/3, to appear in J. Geom.
Phys.}

Gauging world-sheet parity is similar to the usual orbifold construction,
gauging an internal symmetry of the world-sheet theory \ori.  One
difference is that there is no direct analog of the twisted states, because
the Klein bottle does not have the modular transformation $\tau \to
-1/\tau$.  Perturbative unitarity of an orbifold theory {\it requires} the
twisted states, but the unoriented theory is perturbatively unitary without
additional states.  There are however some senses in which open
strings can be regarded as twisted states under world-sheet parity
\ori; we will return to this later.

\newsec{Lecture II: {\sl $T$-Duality}}

\subsec{Self Duality of Closed Strings}

For closed strings, let us first study the zero modes. We have
\eqn\chave{X^\mu (z , \bar{z}) \sim-i\sqrt{\ap\over2}
(\alpha^\mu_0+{\tilde\alpha}^\mu_0)\sigma^2+\sqrt{\ap\over2}
(\alpha^\mu_0-{\tilde\alpha}^\mu_0)\sigma^1 + \cdots .}
Noether's theorem gives the spacetime momentum of a string as
\eqn\noeth{p^\mu =
{1\over{\sqrt{2\ap}}}(\alpha^\mu_0 + {\tilde\alpha}^\mu_0),}
while under $\sigma^1\sim\sigma^1+2\pi$,
$X^\mu(z,\zb)$
changes by $2\pi\sqrt{(\ap/2)}(\alpha^\mu_0-{\tilde\alpha}^\mu_0).$
For a non-compact spatial direction $\mu$, $X^\mu(z,\zb)$ is single-valued,
and so
\eqn\so{\alpha^\mu_0={\tilde\alpha}^\mu_0=\sqrt{\ap\over2}p^\mu.}
Since vertex operators must leave the space~\so\ invariant, only the sum
$x^\mu + \tilde x^\mu$ may appear.

For a compact direction of radius $R$, say $\mu=25$, $X^{25}$ has period
$2\pi R$. The momentum $p^{25}$ can take the values $n/R$.
Now, under $\sigma^1\sim\sigma^1+2\pi$, $X^{25}(z,\zb)$
can change by $2\pi wR$, which means that
\eqn\relations{\eqalign{\alpha^{25}_0+{\tilde\alpha}^{25}_0=&
{2n\over R}\sqrt{\ap\over2}\cr
\alpha^{25}_0-{\tilde\alpha}^{25}_0=&\sqrt{2\over\ap}wR.}}
and so
\eqn\final{\eqalign{\alpha^{25}_0=\left({n\over R}+{wR\over\ap}\right)
\sqrt{\ap\over2}\cr
{\tilde\alpha}^{25}_0=
\left({n\over R}-{wR\over\ap}\right)\sqrt{\ap\over2}}}

Turning to the mass spectrum, we have
\eqn\masses{\eqalign{M^2=-p^\mu p_\mu&=
{2\over\ap}(\alpha_0^{25})^2+{4\over\ap}
({ L}-1)\cr
&={2\over\ap}({\tilde\alpha}_0^{25})^2+{4\over\ap}
( {\bar L}-1).}}
Here $\mu$ runs only over the non-compact dimensions, $ L$ is the
total level of the left-moving excitations, and $ \bar L$ the total
level of the right-moving excitations.
The mass spectra of the theories at radius $R$ and $\ap/ R$ are identical
with the winding and Kaluza-Klein modes interchanged
$n \leftrightarrow w$ \ref\tduspec{
K. Kikkawa and M. Yamanaka, Phys. Lett. {\bf B149} (1984) 357;\hfil\break
N. Sakai and I. Senda, Prog. Theor. Phys. {\bf 75} (1986) 692.},
which takes
\eqn\tdual{\eqalign{\alpha^{25}_0&\rightarrow\alpha^{25}_0\cr
{\tilde\alpha}_0^{25}&\rightarrow-{\tilde\alpha}_0^{25} .}}
The interactions are identical as well \ref\nairet
{V.P. Nair, A. Shapere, A. Strominger, and F. Wilczek,
Nucl. Phys. {\bf B287} (1987) 402.
}.  Write the radius-$R$
theory in terms of
\eqn\xd{X^{\prime25}(z,\zb)=X^{25}(z)-X^{25}(\zb)\ .}
The energy-momentum tensor and OPE and therefore all of the correlation
functions are invariant under this rewriting.  The only change is that the
zero mode spectrum in the new variable is that of the $\ap/R$ theory.

The $T$-duality is therefore an exact symmetry of
perturbative closed string theory.  Note that it can be regarded as a
spacetime parity transformation acting only on the right-moving degrees of
freedom.  We will denote this transformation as $T_{25}$, where $T_{\mu_1
\cdots \mu_k}$ refers to the corresponding transformation on $X^{\mu_1}
\ldots X^{\mu_k}$.

This duality transformation is in fact an {\it exact} symmetry of closed
string theory~\ref\dhs{M. Dine, P. Huet, and N. Seiberg, Nucl. Phys. {\bf
B322} (1989) 301.}. To see why, recall
the  appearance of an $SU(2)_L$$\times$$SU(2)_R$ extended gauge symmetry
at the self-dual radius. Additional left- and right-moving
currents are present at this radius in the massless
spectrum, $\partial X^{25}(z)$, $\exp(\pm 2iX^{25}(z)/\sqrt\ap)$ for
$SU(2)_L$, and ${\bar\partial}X^{25}(\zb)$,
$\exp(\pm 2iX^{25}(\zb)/\sqrt\ap)$ for $SU(2)_R$.
The marginal operator for the change of radius, $\partial X^{25}
{\bar\partial} X^{25}$, transforms as a $({\bf 3},{\bf 3})$, so a rotation
by $\pi$ in one of the $SU(2)$'s transforms it into minus itself.  The
transformation $R \to \ap/R$ is therefore a ${\bf Z}_2$ subgroup of
the $SU(2) \times SU(2)$.  We may not know what non-perturbative string
theory is, but it is a fairly safe bet that it does not violate spacetime
gauge symmetries explicitly, else the theory could not be
consistent.  Note that the $Z_2$ is already spontaneously broken,
away from the self-dual radius.

It is important to note that $T$-duality acts nontrivially on the dilaton.
By the usual dimensional reduction, the effective 25-dimensional coupling is
$e^{\phi} R^{-1/2}$.  Duality requires this to be equal to
$e^{\phi'} R'^{-1/2}$, hence
\eqn\dilt{
e^{\phi'} = e^{\phi} R^{-1} {\ap}^{1/2}
}

\subsec{Open Strings and Dirichlet-Branes}

Let us rewrite the open string mode expansion for the compact direction as
follows:
\eqn\openmodes{\eqalign{X^{25}(z)={x^{25}\over2}
+C-i\ap p^{25}\ln z+i\sqrt{\ap\over2}\sum_{m\neq0}{\alpha^\mu_m\over mz^m},
\cr
X^{25}(\zb)={x^{25}\over2}-C-i\ap p^{25}\ln
\zb+i\sqrt{\ap\over2}\sum_{m\neq0} {\alpha^\mu_m\over m\zb^m}.}}
Then $X^{25}(z,\zb)=X^{25}(z)+X^{25}(\zb)$ is the usual open string coordinate.
Again rewrite the theory in terms of
\eqn\ends{\eqalign{X^{\prime25}(z,\zb)=X^{25}(z)-X^{25}(\zb)&=2C-i\ap p^{25}
\ln(z/\zb)
+({\rm oscillators})\cr
&=2C+2\ap p^{25}\sigma^1
+({\rm oscillators})\cr
&=2C+2\ap {n\over R}\sigma^1 +({\rm oscillators}).}}
\nref\hgdual{
P. Horava, Phys. Lett. {\bf B231} (1989) 251;\hfil\break
M. B. Green, Phys. Lett. {\bf B266} (1991) 325. }
The oscillator terms vanish at the endpoints $\sigma^1 = 0, \pi$.
Notice that there is no dependence on $\sigma^2$ in the zero modes. Therefore
{\it  the endpoints of the string do not move in the $X^{25}$ direction}.  We
could also see this directly, from the boundary condition $0 = \partial_n
X^{25} = \partial_t X^{\prime 25}$ \refs{\dbranes,\hgdual}. At the ends,
\eqn\wherends{\eqalign{
\sigma^1=0:\qquad X^{\prime25}=&2C;\cr
\sigma^1=\pi:\qquad X^{\prime25}=&2C+2\pi\ap p^{25}\cr
=&2C+2\pi nR^\prime.
}}

This means that in the dual theory (with radius $R^\prime=\ap/R$)
the ends of the open strings are located for all time  at position
$X^{\prime25}=2C$. They can wind $n$ times around the spacetime circle,
and they are free
to move in the other directions:

\vskip1.5cm
\hskip3.5cm\epsfxsize=2.0in\epsfbox{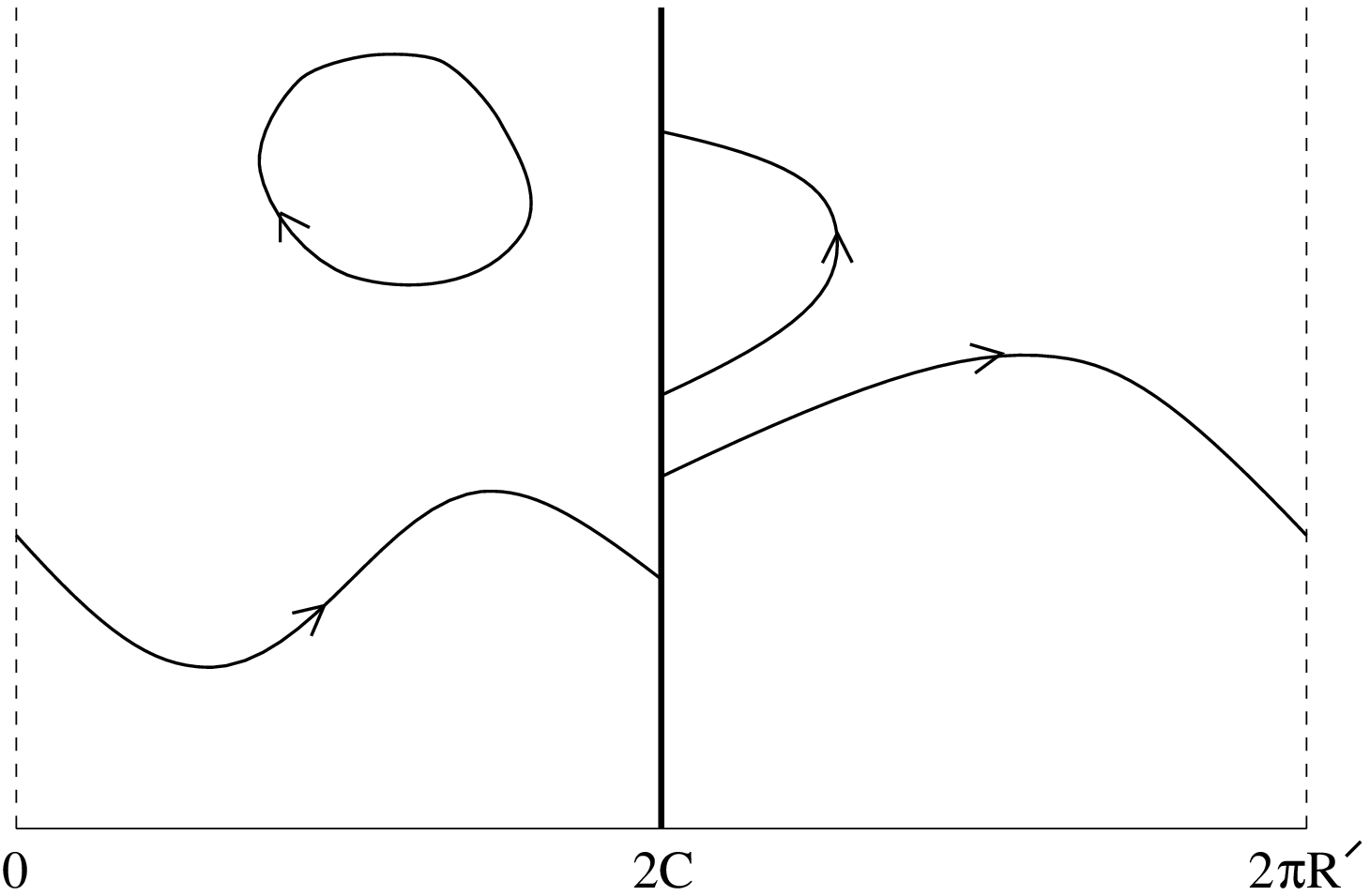}
\vskip0.4cm

In the above diagram, the vertical direction represents the other 24 spatial
directions.
The string endpoints are constrained to lie on the 24-dimensional
hypersurface $X^{\prime25}=2C$; in the present case we could define
$X^{\prime 25}$ such that $C \to 0$, but later $C$ will be necessary.
We will see that this hypersurface is a dynamical object, a membrane, hence
called a Dirichlet-brane (D-brane).

\subsec{Chan-Paton Factors and Multiple D-Branes}

Now we study the effect of
Chan-Paton factors \joeone.
Consider the case of $U(N)$, the oriented open string. In compactifying the
$X^{25}$ direction, we can include a Wilson line $A^{25}={\rm
diag}\{\theta_1,\theta_2,\ldots,\theta_N\}/2\pi R= \partial_{25}\Lambda$,
generically breaking $U(N) \to U(1)^N$.  Locally this is pure gauge,
$\Lambda=(X^{25}/2\pi R){\rm diag}\{\theta_1,\theta_2,\ldots,\theta_N\}$,
but because $X^{25}$ is periodic, $A^{25}$ has
non-trivial holonomy as we go around the circle:
\eqn\holonomy{\Lambda(2\pi R)=\Lambda(0)+{\rm diag}\{\theta_1,\theta_2,
\ldots,\theta_N\}.} We can make a gauge transformation to remove this, but
states which are charged under $U(N)$ pick up a phase in going around the
compact dimension, $|ij\rangle$
being multiplied by $\exp(i[\theta_j-\theta_i])$.  So
\eqn\mom{p^{25}={n\over
R}+{\theta_j-\theta_i\over 2\pi R}} giving
\eqn\zeromodes{\eqalign{X^{\prime25}(z,\zb)=2C&+2\ap p^{25}\sigma^1\cr
=2C&+2\ap\left({n\over R}+{\theta_j-\theta_i\over 2\pi R}\right)\sigma^1.}}
Now we can deduce the value of $C$.  The position of the left endpoint
should depend only on its Chan-Paton degree of freedom $i$ and similarly
for the right endpoint $j$ (else an interaction at one end would have
an instantaneous effect at the other).  If we set $2C=\theta_i R^\prime$ then
indeed
\eqn\wherends{\eqalign{
\sigma^1=0:\qquad X^{\prime25}=&\theta_i R^\prime;\cr
\sigma^1=\pi:\qquad X^{\prime25}=&2\pi nR^\prime+\theta_j R^\prime.}}
That is, an endpoint in state $i$ is located on a D-brane at $\theta_i R'$,
modulo the periodicity of the dual spacetime.

\vskip1.5cm
\hskip3.5cm\epsfxsize=2.0in\epsfbox{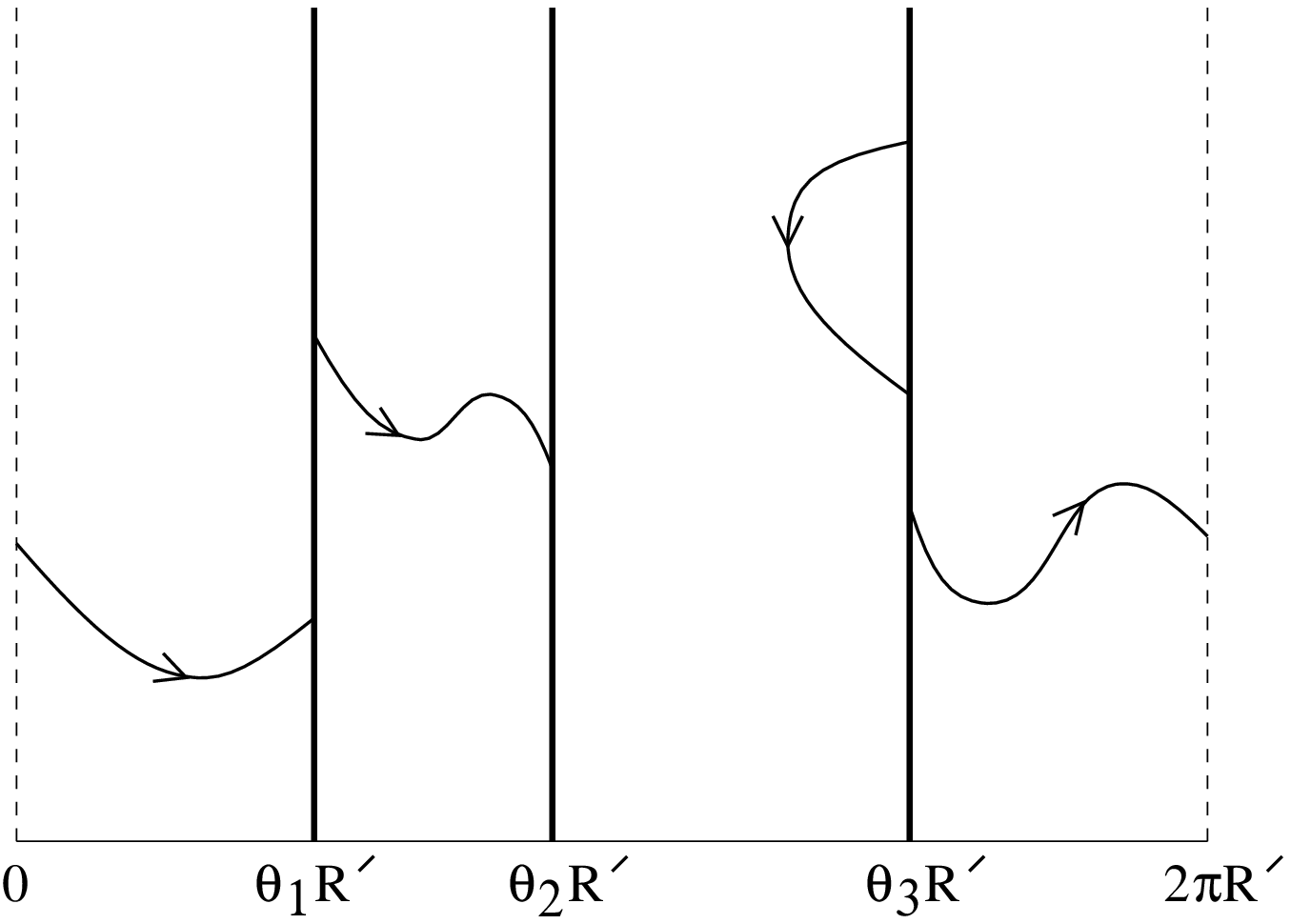}
\vskip0.4cm

Illustrated above are three D-branes with various strings wound between
them.

\subsec{D-Brane Dynamics}

Let us first note that this whole picture goes through if several
coordinates $X^m=\{X^{25},X^{24},\ldots,X^{p+1}\}$ are periodic, and we
rewrite the periodic dimensions in terms of the dual coordinate.  The open
string endpoints are then confined to $N$ $(p+1)$-dimensional
hyperplanes.  The Neumann conditions on the world sheet, $\partial_n
X^m(\sigma^1,\sigma^2)=0$ have become Dirichlet conditions $\partial_t
X^{\prime m}(\sigma^1,\sigma^2)=0$ for the dual coordinates.  The
$(p+1)$-dimensional hypersurface is the world-volume of a $p$-dimensional
extended object called a `Dirichlet $p$-brane', or `D-brane' for
short.\foot{In this terminology, the original Type I theory contains $N$
25-branes. A 25-brane fills space, so the string endpoint can be anywhere:
it just corresponds to an ordinary Chan-Paton factor.}

It is natural to expect that the D-brane is dynamical \dbranes.
Closed strings can interact with the
D-branes (indirectly via open strings), and so the
D-branes feel the effects of gravity in
the closed string massless sector.  We would therefore expect that they
can fluctuate in shape and position as  dynamical objects.
We can see this by looking at the massless spectrum of our theory, interpreted
in the dual coordinates.

Returning to our illustration with a Dirichlet 24-brane, let us look at the
mass spectrum. We have $M^2=(p^{25})^2+{1\over\ap}({ L}-1)$
which gives, for multiple
branes:
\eqn\masses{M^2=\left\{ {[2\pi n+(\theta_i-\theta_j)]R^\prime\over2\pi\ap}
\right\}^2 +{1\over\ap}(L-1).} Note that $[2\pi
n+(\theta_i-\theta_j)]R^\prime$ is the minimum length of the string.  We see
that the massless states arise for non-winding open strings whose end points
are on the same D-brane, as the string tension contributes an energy to a
stretched string.  We have therefore the massless states:
\item{$\qquad\bullet$}{$\alpha^{\mu}_{-1}|{ k},ii\rangle$, vertex operator
$\propto \partial_t X^\mu$. This is the
gauge field in the directions transverse to the D-brane, with $p+1$
components.}
\item{$\qquad\bullet$}{$\alpha^{25}_{-1}|{ k},ii\rangle$, vertex operator
$\propto \partial_t X^{25}=\partial_n X^{\prime25}$.
This is the gauge field in the compact direction of the original theory,
which became the position of the D-brane in the dual theory.  We considered
a classical background for this field, but the string quanta in this state,
which are built into string perturbation theory, correspond to
transverse fluctuations of the D-brane shape.  The relation here is that
same as that between the classical background metric and the graviton states
of the string.}
The world-brane theory thus consists of a $U(1)$ vector field plus $25-p$
world-brane scalars describing the fluctuations.


It is interesting to look at the $U(N)$ symmetry breaking in the dual
picture.  When no D-branes coincide, there is just one massless vector
each, or $U(1)^N$ in all, the generic unbroken group.  If $m$
D-branes coincide, there are new massless states because
(non-winding) strings which are stretched between these branes
can achieve vanishing length.  Thus, there are $m^2$ vectors, forming
the adjoint of a $U(m)$ gauge group. This coincident position
corresponds to  $\theta_1=\theta_2=\cdots=\theta_m$ for some subset of the
original $\{\theta\}$, so in the dual theory the Wilson line
left a $U(m)$ subgroup unbroken.
\nref\witbound{E. Witten, {\it Bound States of Strings and $p$-Branes,}
preprint
IASSNS-HEP-95-83, hep-th/9510135.}
\nref\bound{A. Sen, {\it A Note on Marginally Stable Bound
States in Type II String Theory,} preprint
MRI-PHY-23-95, hep-th/9510229;
{\it $U$ Duality and Intersecting D-Branes,} preprint
MRI-PHY-27-95, hep-th/9511026;\hfil\break
C. Vafa, {\it Gas of D-Branes and Hagedorn Density of BPS States,} preprint
HUTP-95-A042, hep-th/9511088; {\it Instantons on D-Branes,}
preprint HUTP-95-A049, hep-th/9512078.}
At the same time, there
appears the set of $m^2$ massless scalars: the $m$ positions are promoted
to a matrix.  This is curious and hard to visualize, but has proven to play
an important role in the dynamics of D-branes~\refs{\witbound,\bound}.
As one consequence, consider the figure, which shows two singly wound
strings and one doubly  wound string on a compact dimension.

\vskip0.75cm
\hskip4.5cm\epsfxsize=1.5in\epsfbox{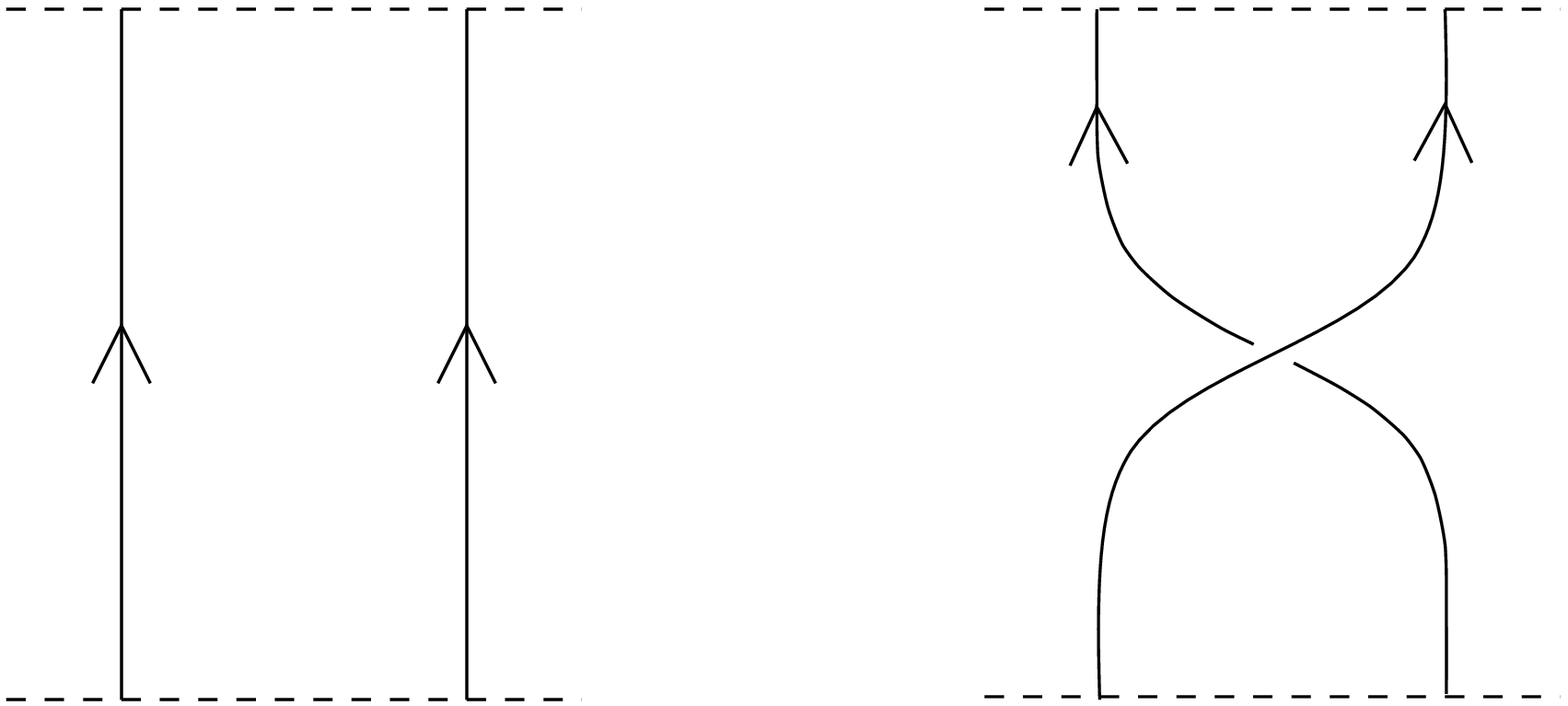}
\vskip0.5cm

For fundamental strings these are distinct.  For D-branes, however, the
integral over the D-brane $U(2)$ group will include an integral over
the holonomy in going around the compact dimension.  Since
this acts on the two D-branes, the two parts of the figure just
represent holonomies
\eqn\holos{
\pmatrix{1&0\cr0&1}, \qquad \pmatrix{0&1 \cr 1&0} }
respectively, and are continuously connected through configurations
that cannot be drawn!\foot
{M. Douglas (seminar at ITP) has made the interesting observation that
one can think of this as an enlargement of the usual statistics group on
$m$ particles from $S_m$ to $U(m)$!} Note that if all
$N$ branes are coincident, we recover the
$U(N)$ gauge symmetry.

This picture seems a bit exotic, and will become more so in the unoriented
theory.  But all we have done is to rewrite the original open string theory
in terms of variables which are more natural in the limit $R <<
\sqrt{\ap}$.  Various puzzling features of the small-radius limit become
clear in the $T$-dual picture.

\subsec{D-Brane Tension}

One can use nonlinear sigma model methods to
find the conditions for conformal invariance of the D-brane CFT.  The
field equations lead to an effective action for a D-brane moving in a
closed string background~\nref\frad{E. S. Fradkin and A. A. Tseytlin, Phys.
Lett. {\bf B163} (1985) 123.}\refs{\frad,\dbranes}
\eqn\branetheory{S=-T_{p}\int d^{p+1}\sigma\left\{ e^{-\phi}\Tr\, {\rm
det}^{1\over2}\left({\tilde G}_{\mu\nu}+{\tilde B}_{\mu\nu}+
2\pi \ap F_{\mu\nu}
\right) \right\},}
where ${\tilde G}_{\mu\nu}$ and ${\tilde B}_{\mu\nu}$ are the induced
fields on the world brane, $F_{\mu\nu}$ is the open string
$U(1)$ gauge field strength, and the trace is over the Chan-Paton degrees
of freedom.  The tension is of
order $e^{-\phi} = g^{-1}$, because the D-brane is invisible to
amplitudes on the sphere and contributes first at disk order.  The action
for the field strength can be understood from $T$-duality.  Start with the
original Type I theory but now with a background $A^{25}(X)$ depending on
$X^\mu$ for $\mu \neq 25$.  By the earlier construction, the dual is now a
curved D-brane, $X^{\prime 25}(X) = 2\pi\ap A^{25}$.  The
Dirac-Born-Infeld action then gives the area of this D-brane.  Because
$B_{\mu\nu}$ appears, the antisymmetric tensor gauge invariance must act
on the photon as well,
\eqn\gaugeinv{\eqalign{{\tilde
B}_{\mu\nu}&\to{\tilde B}_{\mu\nu}+\partial_\mu
\chi_\nu-\partial_\nu\chi_\mu\cr A_\mu&\to A_\mu-\chi_\mu .}}
At the
world-sheet level this occurs because a surface term from the variation of
$B_{\mu\nu}$ must be canceled by variation of the open string vector.

It is instructive to compute the D-brane tension $T_{p}$. As noted above,
this is proportional to $g^{-1}$. It arises
from considering the disk tadpole, where in this case the disk is trapped
with its boundary on a D-brane:

\vskip0.75cm
\hskip5cm\epsfxsize=1.5in\epsfbox{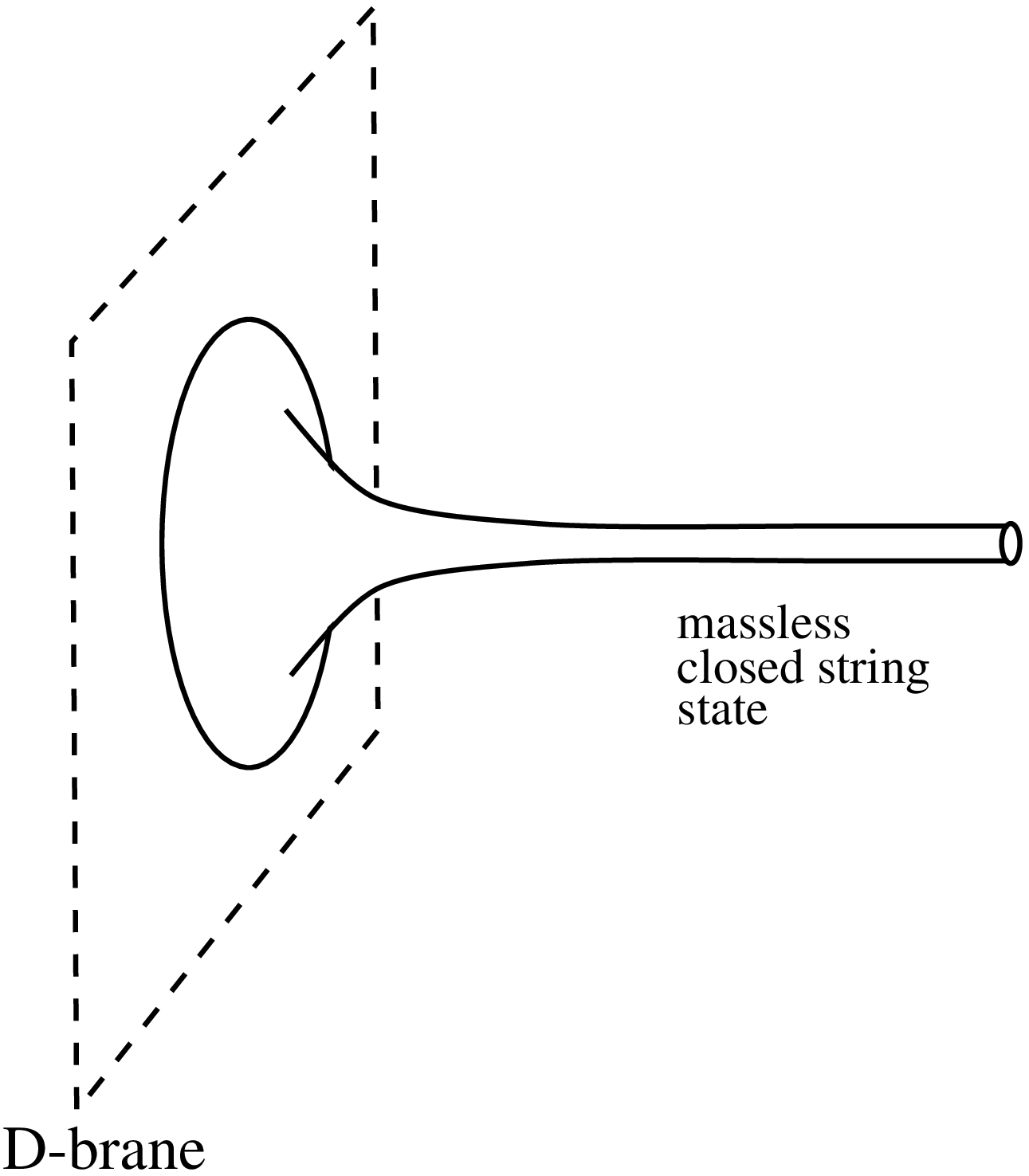}

One could obtain the tension by calculating the disk with a graviton
vertex operator, but it is easier to proceed as follows.
Consider two
parallel Dirichlet $p$-branes at positions $X^{\prime\mu}=0$ and
$X^{\prime\mu}=Y^\mu$. These two objects can feel each other's presence by
exchanging closed strings,

\vskip0.5cm
\hskip4cm\epsfxsize=2.0in\epsfbox{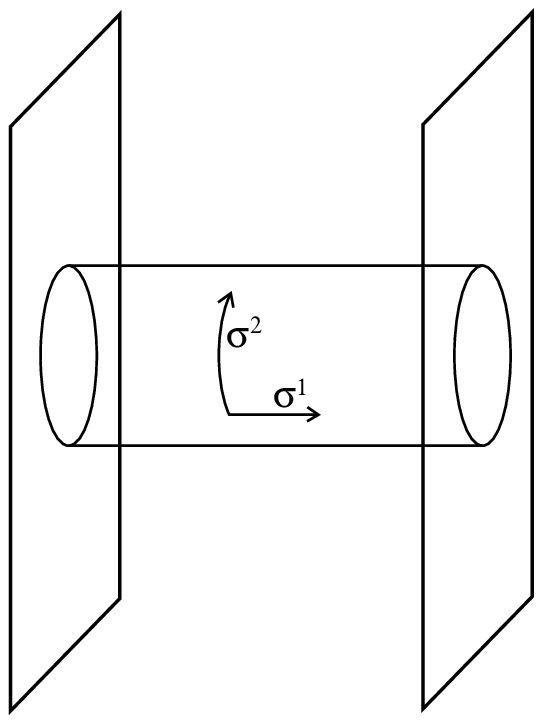}
\vskip0.5cm

This string graph is an annulus, with no vertex operators.
It is therefore easily calculated.  The poles from graviton and dilaton
exchange then give the coupling of closed string states to the D-brane,
that is, $T_p$.

Parameterize the world-sheet as $(\sigma^1,\sigma^2)$ where
$\sigma^1$  runs from $0$ to $\pi$, and $\sigma^2$ is a periodic
coordinate running  from $0$ to $2\pi t$.
\nref\chan{ C. Lovelace, Phys. Lett. {\bf B34} (1971) 500;\hfil\break
L. Clavelli and J. Shapiro, Nucl. Phys. {\bf B57} (1973) 490;\hfil\break
M. Ademollo, R. D' Auria, F. Gliozzi, E. Napolitano, S. Sciuto, and P. di
Vecchia, Nucl. Phys. {\bf B94} (1975) 221;\hfil\break
C. G. Callan, C. Lovelace, C. R. Nappi, and S. A. Yost,
Nucl. Phys. {\bf B293} (1987) 83.}
\nref\rrex{
J. Polchinski and Y. Cai, Nucl. Phys. {\bf B296} (1988) 91;\hfil\break
C. G. Callan, C. Lovelace, C. R. Nappi and S.A. Yost,
Nucl. Phys. {\bf B308} (1988) 221. }
This vacuum graph has the single
modulus $t$, running from $0$ to $\infty$.  If we time-slice
horizontally, so that $\sigma^2$ is world-sheet time, we see two open
strings appearing out of the vacuum and then disappearing, giving the  open
string loop channel.  Time-slicing vertically
instead, so that $\sigma^1$ is time, we see a single closed string
propagating in the tree channel. The world-line of the open string boundary
can be regarded as a vertex connecting the vacuum to the single closed
string, i.e., a one-point closed string vertex.  The two channels are
related by duality
\refs{\chan,\rrex}.

Consider the limit $t$$\to$$0$ of the loop amplitude. This is the
ultra-violet limit for the open string channel, but unlike the torus, there
is no modular group acting to cut off the range of integration.  However,
because of duality, this limit is correctly interpreted as an {\it
infrared} limit. Time-slicing in the vertical direction shows that the
$t$$\to$$0$ limit is dominated by the lowest lying modes in the
closed string spectrum.  In keeping with string folklore, there are no
``ultraviolet limits'' of the moduli space  which could give rise to high
energy divergences.  All divergences in loop amplitudes come from pinching
handles and are controlled by the lightest states, or the long distance
physics.

One loop vacuum amplitudes are given by the Coleman-Weinberg
formula, which amounts to summing the zero point energies of all the
modes \ref\cw{S. Coleman and E. Weinberg, Phys. Rev. {\bf D7} (1973)
1888;\hfil\break J. Polchinski, Comm. Math. Phys. {\bf 104}
(1986) 37.}:
\eqn\vac{{\cal A} = \int_0^\infty {dt \over 2t} \sum_{i,k}
e^{-2\pi \ap t (k^2 + M_i^2)} .}
Here the sum $i$ is over the physical spectrum of the string, equivalent
to the transverse spectrum, and the momentum ${\bf k}$ is in
the $p+1$ extended directions of the D-brane world-sheet.
The mass spectrum is given by
\eqn\mass{M^2={1\over\ap}\left(\sum_{n=1}^\infty
n{\alpha}^i_{-n}\alpha_n^i-1\right)+{Y \cdot Y\over4\pi^2
\alpha^{\prime 2}}.}
where $Y^m = x^m_1 - x^m_2$ is the separation of the D-branes.
The sums over $N_n^i \equiv {\alpha}^i_{-n}\alpha_n^i$ are as usual
geometric, giving
\eqn\potential{
{\cal A}=2V_{p+1} \int_0^\infty {dt\over 2t} (8\pi^2 \ap t)^{-{(p+1)\over2}}
e^{-Y \cdot Y t/ 2\pi\ap} q^{-2} \prod_{n=1}^\infty
(1-q^{2n})^{-24}}
where $q=e^{-\pi t}$ and the overall factor of 2 is from exchanging the
two ends of the string.  We need the asymptotics as $t \to 0$.  More
generally, define
\eqn\thet{\eqalign{
f_1(q) &= q^{1/12} \prod_{n=1}^{\infty} (1-q^{2n})
\cr
f_2(q) &= \sqrt{2} q^{1/12} \prod_{n=1}^{\infty} (1+q^{2n})
\cr
f_3(q) &= q^{-1/24} \prod_{n=1}^{\infty} (1+q^{2n-1})
\cr
f_4(q) &= q^{-1/24} \prod_{n=1}^{\infty} (1-q^{2n-1})
}}
The asymptotics as $t\to \infty$ are manifest.  The asymptotics as $t\to
0$ are then obtained from the modular transformations
\eqn\mod
{ f_{1}(e^{-{\pi}/{s}}) = \sqrt{s}\,f_{1}(e^{-\pi s}),\quad
f_{3}(e^{-{\pi}/{s}}) = f_{3}(e^{-\pi s}),\quad
f_{2}(e^{-{\pi}/{s}}) = f_{4}(e^{-\pi s}) . }
In the present case
\eqn\amplitude{
{\cal A}=
2 V_{p+1} \int_0^\infty {dt\over 2t} (8\pi^2 \ap t)^{-{(p+1)\over2}}
e^{-Y \cdot Y t/ 2\pi\ap} t^{12} \left( e^{2\pi/t} + 24 + \ldots
\right).}
The leading divergence is from the tachyon and is an uninteresting bosonic
artifact.  The massless pole, from the second term, is
\eqn\pole{\eqalign{
{\cal A} &\sim V_{p+1}{24\over 2^{12}} (4\pi^2\ap)^{11-p}
\pi^{(p-23)/2} \Gamma((23-p)/2) |Y|^{p-23} \cr
& = V_{p+1} {3 \pi\over 2^{7}} (4\pi^2\ap)^{11-p}
G_{25-p}(Y^2)}}
where $G_D(Y^2)$ is the massless scalar Green's function in $D$ dimensions.

This can be compared with a field theory calculation, the exchange of
graviton plus dilaton between a pair of D-branes, with the bulk
action~\lowlimit\ and the coupling~\branetheory\ to the D-brane.
This is a bit of effort because the graviton and dilaton mix, but in the
end one finds
\eqn\fieldpole{
{\cal A} \sim {D-2 \over 4} V_{p+1} T_p^2 G_{25-p}(Y^2)
}
so
\eqn\tens{ T_p = {\sqrt{\pi}\over 16} (4\pi^2\ap)^{(11-p)/2}. }
The units are obscured because we are working with dimensionless
$\kappa = e^{\phi}$.  The physical tension is $\tau_p
= e^{-\phi}T_p = T_p/\kappa$, which is dimensionally correct.

As one application, consider $N$ 25-branes, which is just an ordinary
$N$-valued Chan-Paton factor.  Expanding the 25-brane
Lagrangian~\branetheory\ to second order in the gauge field gives
\eqn\gaugec{{T_{25} \over 4} (2\pi\ap)^2 e^{-\phi}\Tr F_{\mu\nu}F^{\mu\nu},}
with the trace in the fundamental representation of $U(N)$.  This gives the
precise numerical relation between the open and closed string
couplings \ref\occoup{J. A. Shapiro and C. B. Thorn, Phys. Rev. {\bf D36}
(1987) 432;\hfil\break J. Dai and J. Polchinski, Phys. Lett. {\bf B220}
(1989) 387.}.

The asymptotics~\amplitude\ have an obvious interpretation in terms of
a sum over closed string states exchanged between the two D-branes.
One can write the cylinder path integral in Hilbert space formalism
treating
$\sigma_1$ rather than $\sigma_2$ as time.  It then has the form
\eqn\bsform{
\langle B | e^{-(L_0 + \tilde L_0)\pi/t} | B \rangle}
where the {\it
boundary state} $|B\rangle$ is the closed string state created by the
boundary loop.  We will not have time to develop
this formalism but it is useful in finding the couplings between
closed and open strings \refs{\chan,\rrex}.

\subsec{Unoriented Strings and Orientifolds.}

For closed strings, the original coordinate is
$X^m(z,\zb)=X^m(z)+X^m(\zb)$ and the dual is $X^{\prime
m}(z,\zb)=X^m(z)-X^m(\zb)$.  The action of world sheet parity reversal is
to exchange $X^\mu(z)$ and $X^\mu(\zb)$. In terms of the dual coordinate
this is
\eqn\omegadual{
X^{\prime m}(z) \leftrightarrow -X^{\prime m}(\zb),
}
which is the product of a world-sheet and spacetime parity operation.
In the unoriented theory, strings are invariant under the action of
$\Omega$.  Separate the string wavefunction into its internal part and
its dependence on the center of mass $x^m$, and take the internal
wavefunction to be an eigenstate of $\Omega$.  The projection then
determines the wavefunction at $-x^m$ to be the same as at $x^m$, up to a
sign.  This is the same as the orbifold construction, the only difference
being that the internal part includes a world-sheet parity reversal; thus
we will call it an {\it orientifold} \refs{\ori,\dbranes}.
The compact spacetime is
effectively the orbifold $T^{25-p}/Z_2$.  For the case of a single compact
dimension, for example, spacetime is the line segment $0 \leq x^{25} \leq
\pi R'$,  with orientifold fixed planes at the ends.
It should be noted that orientifold planes are not dynamical.  Unlike the
case of D-branes, there are no string modes tied to the orientifold plane
to represent fluctuations in its shape.

In the case of open strings, the situation is similar.
Let us focus for convenience on a single compact dimension.  Again
there is one orientifold fixed plane at $0$ and another at $\pi R^\prime$.
Introducing $SO(N)$ Chan-Paton factors, a Wilson line can be brought to
the form
\eqn\wilson{\pmatrix{0&i\theta_1&0&0&\cdots&\cr
-i\theta_1&0&0&0&\cdots&\cr 0&0&0&i\theta_2&\cdots&\cr
0&0&-i\theta_2&0&\cdots&\cr
\vdots&\vdots&\vdots&\vdots&\ddots}}
or equivalently
\eqn\dor{{\rm diag}
\{\theta_1,-\theta_1,\theta_2,-\theta_2,\cdots,\theta_{N/2},-\theta_{N/2}\}.
}
Thus in the dual picture there are
${1\over 2}N$ D-branes on the line segment $0\leq X^{\prime25}<\pi
R^\prime$, and ${1\over 2}N$ at their image points under the orientifold
identification. 

\vskip0.75cm
\hskip2.5cm\epsfxsize=3.0in\epsfbox{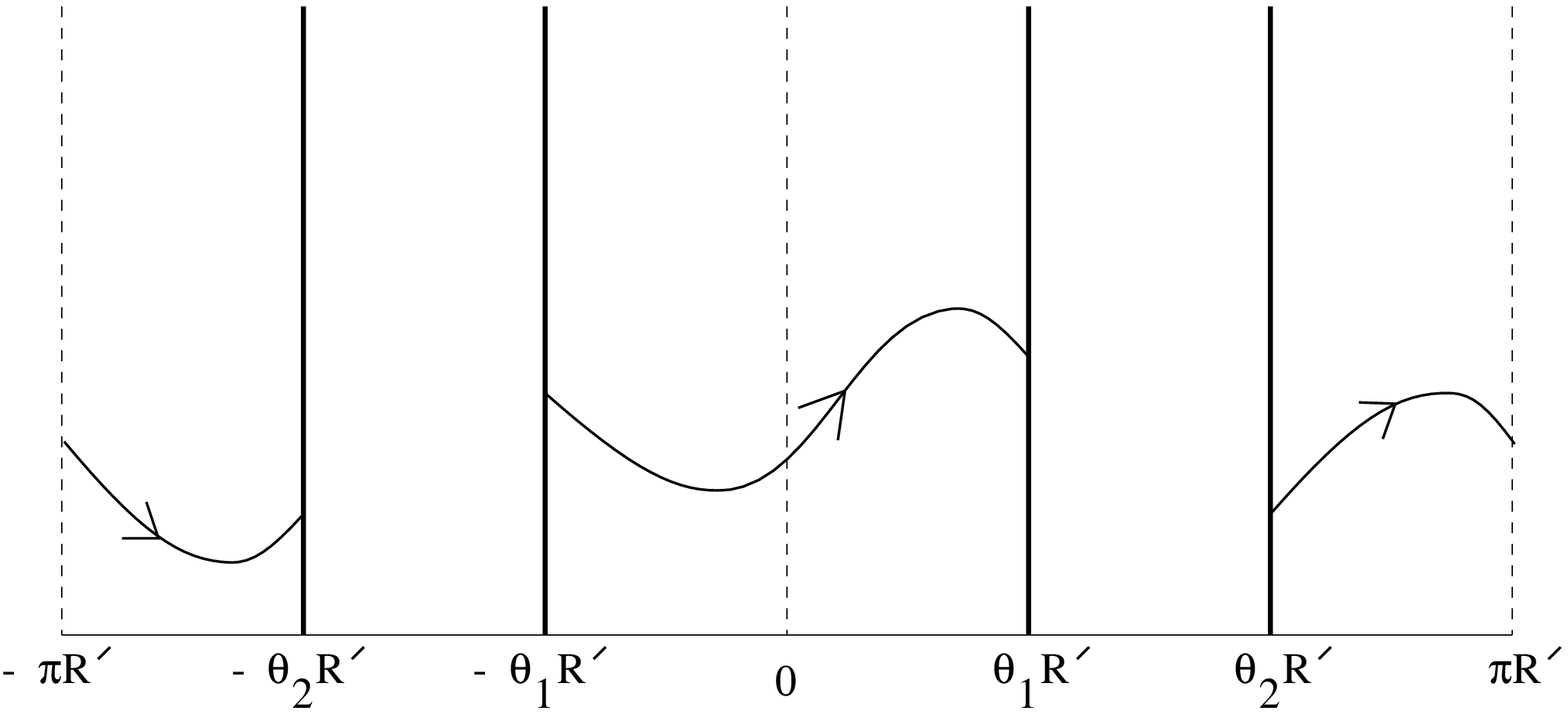}

Strings can stretch between D-branes and their images as shown.
The generic gauge group is again $U(1)^{N/2}$.  As in the oriented case,
if $m$ D-branes are coincident there is a $U(m)$ gauge group.
Now if the $m$ D-branes in addition lie at one of the fixed
planes, then strings stretching between one of these branes and one of the
image branes also become massless and we
have the right spectrum of additional states to fill out $SO(2m)$. The
maximal $SO(N)$ is restored if all of the branes are coincident at a
single orientifold plane.  Note that this maximally symmetric case is
asymmetric among the fixed planes, a fact that will play an important role
later.  Similar considerations apply to $USp(N)$.

The orientifold plane, like the D-brane, couples to the dilaton and metric.
The amplitude is the same as in the previous section, but with $RP^2$
in place of the disk; that is, a crosscap replaces the boundary loop.  The
orientifold identifies $X^m$ with $-X^m$ at the opposite point on the
crosscap, so the crosscap is localized near one of the orientifold fixed
planes.  Again the easiest way to calculate this is via vacuum graphs, the
cylinder with one or both boundary loops replaced by crosscaps.  These
give the M\"obius strip and Klein bottle, respectively:

\hskip3.5cm\epsfxsize=2.0in\epsfbox{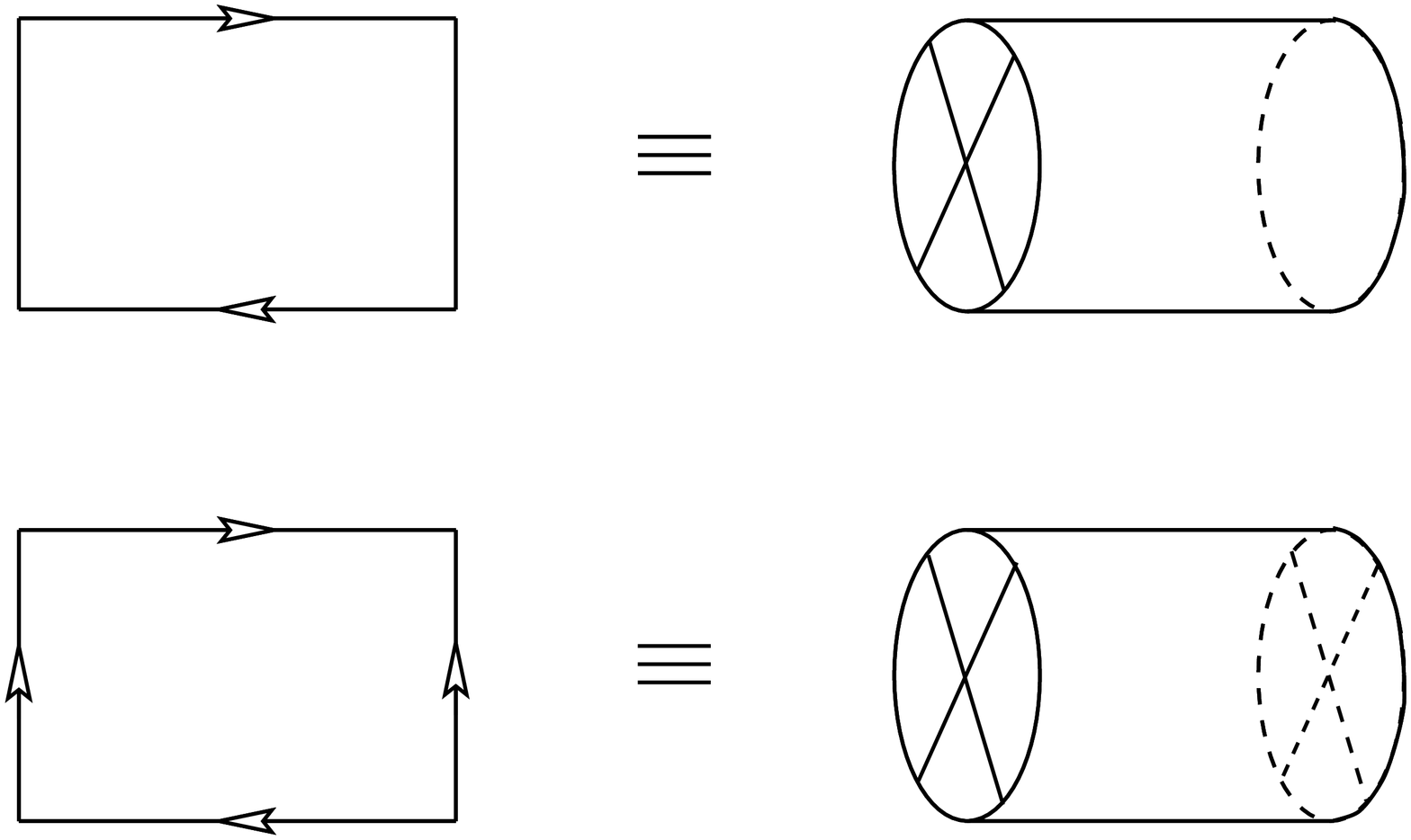}

To understand this, consider the figure below, which shows two copies of
the fundamental region for the M\"obius strip:

\vskip1.0cm
\hskip5.0cm\epsfxsize=1.0in\epsfbox{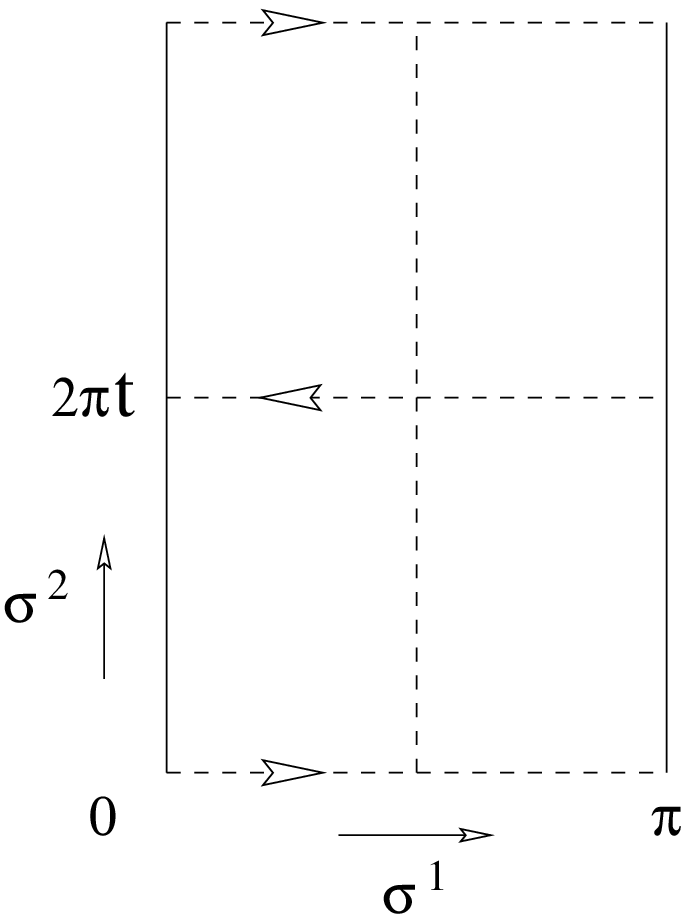}
\vskip0.4cm

The lower half is identified with the reflection of the upper, and the
edges $\sigma^1 = 0, \pi$ are boundaries.  Taking the lower half as the
fundamental region gives the familiar representation of the M\"obius strip
as a strip of length $2\pi t$, with ends twisted and glued.  Taking instead
the left half of the figure, the line $\sigma^1 = 0$ is a boundary loop
while the line $\sigma^1 = \pi/2$ is identified with itself under a shift
$\sigma^2 \to \sigma^2 + 2\pi t$ plus reflection of $\sigma^1$: it is a
crosscap.  The same construction applies to the Klein bottle, with the
right and left edges now identified.

The M\"obius strip is now given by the vacuum amplitude
\eqn\vacmob{{\cal A}_{\rm M} = \int_0^\infty {dt \over 2t} \sum_{i,k}
{\Omega_i\over 2} e^{-2\pi \ap t (k^2 + M_i^2)} ,}
where $\Omega_i$ is the $\Omega$ eigenvalue of state $i$.  The oscillator
contribution to $\Omega_i$ is $(-1)^L$ from eq.~\have.\foot
{In the compact directions there are two additional signs that cancel: the
world-sheet parity contributes an extra minus sign in the directions
with Dirichlet boundary conditions, and the spacetime part an additional
sign.}
For
the $SO(N)$ open string the Chan-Paton factors have ${1\over 2}N(N+1)$ even
states and ${1\over 2}N(N-1)$ odd for a net of $+N$. For $USp(N)$ these
numbers are reversed, for a net of $-N$.  Focus on a D-brane and its image,
which correspondingly contribute $\pm 2$.  The diagonal elements, which
contribute to the trace, are those where one end is on the D-brane and one
on its image.  The total separation is then $Y^m = 2x^m + 2\pi n^m R'$,
corresponding to a fixed plane (or its period image) at $x^m = \pi n^m
R'$.  Taking into account these factors, the $n^m = 0$ term is
\eqn\potential{
{\cal A}_{\rm M}= \pm V_{p+1} \int_0^\infty {dt\over 2t} (8\pi^2 \ap
t)^{-{(p+1)\over2}} e^{-2x \cdot x t/ \pi\ap}
\left[ q^{-2} \prod_{k=1}^\infty (1+q^{4k-2})^{-24} (1-q^{4k})^{-24} \right]
}
The factor in braces $[\ ]$ is
\eqn\fs{\eqalign{
f_3(q^2)^{-24} f_1(q^2)^{-24} &= (2t)^{12}
f_3(e^{-\pi/2t})^{-24} f_1(e^{-\pi/2t})^{-24} \cr
&= (2t)^{12} \left(e^{\pi/2t} - 24 + \ldots \right).
}}
One thus finds a pole
\eqn\mpole{\mp 2^{p - 12} V_{p+1} {3 \pi\over 2^{7}}  (4\pi^2\ap)^{11-p}
G_{25-p}(X^2).
}
This is to be compared with ${1\over2} (D-2) T_p T'_p G_{25-p}(Y^2)$, where
$T'_p$ is the fixed-plane tension; a factor of 2 as compared to the
earlier field theory calculation~\fieldpole\ comes from the spacetime
boundary.  Thus the fixed-plane and D-brane tensions are related
\eqn\orten{
T'_p = \mp 2^{p - 13} T_p.
}
A similar calculation with the Klein bottle gives a result proportional
to $T_p^{\prime 2}$.

Noting that there are $2^{25 - p}$ fixed planes, the total fixed-plane
source is $\mp 2^{12} T_p$.  The fixed-plane and D-brane sources
cancel for the group $SO(2^{13}) = SO(2^{D/2})$ \ref\sobig{M. Douglas and
B. Grinstein, Phys. Lett. {\bf B183} (1987) 552; (E) {\bf 187} (1987) 442. S.
Weinberg, Phys. Lett. {\bf B187} (1987) 278;\hfil\break
N. Marcus and A. Sagnotti, Phys. Lett. {\bf B188} (1987) 58.}.\foot{
This corresponds to
$2^{12}$ D-branes--it would be overcounting to include also their images.
Incidentally, we use the D-brane tension of the oriented theory, because the
local physics away from the fixed planes is oriented; in the unoriented
theory the tension is smaller by
$\sqrt 2$.} For this group the dilaton and graviton tadpoles cancel at order
$g^{-1}$.  This has no special significance in the bosonic string, as the
one loop $g^0$ tadpoles are nonzero and imaginary due to the tachyon
instability, but similar boundary combinatorics
will give a restriction on anomaly free Chan Paton gauge groups in the
superstring.

The M\"obius strip and Klein
bottle, like the cylinder, can be written in terms of the closed string
Hilbert space~\refs{\chan,\rrex}.  Like a boundary loop, the crosscap
can be thought of as creating a closed string in a state
$|C\rangle$.  The two amplitudes are then
\eqn\ccform{
\langle B | e^{-(L_0 + \tilde L_0)\pi/4t} | C \rangle,
\qquad
\langle C | e^{-(L_0 + \tilde L_0)\pi/2t} | C \rangle
}
where the different $t$ dependences in eqs.~\bsform, \ccform\ follow from
the mapping between the two ways of drawing each surface.

This concludes our survey of bosonic open and unoriented strings and their
$T$-dualities.  The final picture is rather exotic, but remember that this
is just the original string theory, rewritten in terms of variables which
display most clearly the physics of the $R \to 0$, $R' \to \infty$, limit.

\newsec{Lecture III: {\sl Superstrings and $T$-Duality}}

\subsec{Open Superstrings}
All of the exotic phenomena that we found in the bosonic string will
appear in the superstring as well, together with some important new
ingredients.  We first review open and unoriented superstrings.

The superstring world-sheet action is
\eqn\action{
S = {{1}\over{4 \pi}} \int_{\cal M} d^2\sigma
  \{ \alpha^{\prime-1} \partial X^{\mu} \bar{\partial} X_{\mu}
     + \psi^{\mu} \bar{\partial} \psi_{\mu} +
       \tilde{\psi}^{\mu}\partial \tilde{\psi}_{\mu} \}
}
where the open string world-sheet is the strip
$0 < \sigma^1 < \pi$, $- \infty <\sigma^2< \infty$.
The condition that the surface term in the equation of
motion vanishes allows two possible Lorentz invariant boundary conditions
on world-sheet fermions:
\eqn\bcf{\eqalign{
{\rm\bf R} \qquad & \psi^{\mu}(0, \sigma^2) =\tilde{\psi^{\mu}} (0,\sigma^2)
 \qquad\psi^{\mu}(\pi ,\sigma^2 )= \tilde{\psi^{\mu}}(\pi,\sigma^2)\cr
  {\rm\bf NS} \qquad & \psi^{\mu}(0, \sigma^2) =-\tilde{\psi^{\mu}}
(0,\sigma^2)
   \qquad\psi^{\mu}(\pi ,\sigma^2 )= \tilde{\psi^{\mu}}(\pi,\sigma^2)\cr
}}
We can always take the boundary condition at one end, say $\sigma^1 =
\pi$, to have a $+$ sign by redefinition of $\tilde\psi$.  The boundary
conditions and equations of motion are conveniently summarized by the {\it
doubling trick}, taking just left-moving (analytic) fields $\psi^\mu$ on
the  range $0$ to $2\pi$ and defining $\tilde\psi^\mu(\sigma^1,\sigma^2)$
to be $\psi^\mu(2\pi - \sigma^1,\sigma^2)$.
These left-moving fields are periodic in the Ramond (R) sector and
antiperiodic in the Neveu-Schwarz (NS).

In the NS sector the fermionic
oscillators are half-integer moded, giving a ground state energy of
\eqn\nszero{ \left (-{{8}\over{24}}\right )+\left
(-{{8}\over{48}}\right )
 = - {1\over 2}
}
from the eight transverse coordinates and eight transverse
fermions.  The ground state is a
Lorentz singlet and has odd
fermion number, $(-1)^F = -1$.  This assignment is necessary in order for
$(-1)^F$ to be multiplicatively conserved.\foot{In the `$-1$ picture'
\ref\fms{D. Friedan, E. Martinec, and S. Shenker, Nucl. Phys. {\bf B271}
(1986) 93.} the matter part of the ground state vertex operator is the
identity but the ghost part has odd fermion number.  In the `0 picture' this
is reversed.} The GSO projection, onto states with even fermion number,
removes the open string tachyon from the superstring spectrum.  Massless
particle states in ten dimensions are classified by their $SO(8)$
representation under Lorentz rotations which leave the momentum invariant.
The lowest lying states in the NS sector are the eight transverse
polarizations of the massless open string photon, $A^{\mu}$,
\eqn\vector{
\psi^{\mu}_{-1/2} |{ k}\rangle, \qquad M^2={{1}\over{\alpha'}} (L - \ha)
}
forming the vector of $SO(8)$.

The fermionic oscillators in the Ramond sector are integer-moded.
In the R sector the ground state energy always vanishes
because the world-sheet bosons and their superconformal partners have the
same moding.\foot{This will remain true later when some bosons are
integer moded and some half-integer.  Note the the R and NS sectors are
always identified by the periodicity properties of the world-sheet
supercurrent.}
The Ramond vacuum is degenerate, since the $\psi^{\mu}_0$ take ground
states into ground states, so the latter form a representation of the
ten-dimensional Dirac matrix algebra
\eqn\dirac{
\{ \psi^{\mu}_0 , \psi^{\nu}_0 \} = \eta^{\mu \nu}
}
The following basis for this representation is often convenient.  Form
the combinations
\eqn\clifford{\eqalign{
d^{\pm}_i &={{1}\over{\sqrt 2}}\left ( \psi^{2i}_0\pm i \psi^{2i+1}_0\right )
 \qquad i=1,\cdots,4 \cr
  d^{\pm}_0 &={{1}\over{\sqrt 2}}\left ( \psi^{1}_0 \mp \psi^{0}_0\right )
}}
In this basis, the Clifford algebra takes the form
\eqn\cli{
\{ d^{+}_i, d^{-}_j \}=\delta_{ij}
}
The $d^{\pm}_i$, $i = 0, \cdots, 4$ act as raising and lowering
operators, generating the 32 Ramond ground states.  Denote these
states
\eqn\bfs{
|s_0,s_1,s_2,s_3,s_4 \rangle = |{\bf s}\rangle
}
where each of the $s_i$ is $\pm\ha$, and where
\eqn\vac{
d^{-}_{i} | -\ha , -\ha , -\ha , -\ha , -\ha \rangle = 0
}
while $d^{+}_i$ raises $s_i$ from $-\ha$ to $\ha$.
The significance of this notation is as follows.  The fermionic part of the
ten-dimensional Lorentz generators is
\eqn\lorentz{
S^{\mu \nu} = - {{i}\over{2}} \sum_{r \in {\bf Z }+\kappa}
  \psi^{[ \mu}_{-r},\psi^{\nu ]}_r
}
where $\kappa$$=$$0$($\ha$) in the R(NS) sector.  The states above are
eigenstates of $S_0 = iS^{01}$, $S_i = S^{2i,2i+1}$, with $s_i$ the
corresponding eigenvalues.  Since the Lorentz generators always flip an
even number of $s_i$, the Dirac representation $\bf 32$ decomposes into a
$\bf 16$ with an even number of $-\ha$'s and $\bf 16'$ with an odd number.

Physical states are annihilated by the zero mode of the supersymmetry
generator, which on the ground states
reduces to $G_0$$=$$p_{\mu}\psi^{\mu}_0$. In the frame $p^0 = p^1$ this
becomes $s_0 = \ha$,
giving a sixteen-fold degeneracy for the {\it physical}
Ramond vacuum.  This is a representation of $SO(8)$ which again
decomposes into ${\bf 8_s}$ with an even number of $-\ha$'s and ${\bf
8_c}$ with an odd number.

The GSO projection keeps one irreducible representation; the two choices,
$\bf 16$ or $\bf 16'$, are physically equivalent, differing only by a
spacetime parity redefinition.
It is useful
to think of the GSO projection in terms of locality of the OPE with the
gravitino vertex operator.  Suppose we take a projection which includes
the operator $e^{-\varphi/2 + i(H_0+H_1+H_2+H_3+H_4)/2}$, where the $H_i$
are the bosonization of $\psi^\mu$.  In
the NS sector this has a branch cut with the ground state vertex operator
$e^{-\varphi}$, accounting for the sign discussed above.  In the R sector
the ghost plus longitudinal part is local, so we have
\eqn\gso{\sum_{i=1}^4 s_i = 0 \pmod 2,}
picking out the ${\bf 8_s}$.

The ground state spectrum is then ${\bf 8}_v \oplus {\bf 8}_s$, a vector
multiplet of $D=10$, $N=1$ spacetime supersymmetry.  Including Chan-Paton
factors gives again a $U(N)$ gauge theory in the oriented theory and
$SO(N)$ or $USp(N)$ in the unoriented.

\subsec{Closed Superstrings}
The closed string spectrum is the product of two copies of the open
string spectrum, with right- and left-moving levels matched.
In the open string the two choices for the GSO projection were
equivalent, but in the closed string there are two inequivalent choices,
taking the same (IIb) or opposite (IIa) projections on the two sides.
These lead to the massless sectors
\eqn\types{\eqalign{
{\rm Type ~ IIa} &\qquad\qquad ({\bf 8_v}\oplus{\bf 8_s}) \otimes
   ({\bf 8_v}\oplus{\bf 8_{c}}) \cr
{\rm Type ~ IIb} &\qquad\qquad({\bf 8_v}\oplus{\bf 8_s}) \otimes
   ({\bf 8_v}\oplus{\bf 8_{s}})
}}
of $SO(8)$.

The various products are as follows.  In the NS-NS sector,
this is
\eqn\prodnsns{
{\bf 8_v} \otimes {\bf 8_v} = \phi \oplus B_{\mu\nu} \oplus G_{\mu\nu}
={\bf 1} \oplus {\bf 28}  \oplus {\bf 35} .}
In the R-R sector, the IIa and IIb spectra are respectively
\eqn\prodrr{\eqalign{
{\bf 8_s} \otimes {\bf 8_c} &= [1] \oplus [3] = {\bf 8_v} \oplus
{\bf 56_t} \cr
{\bf 8_s} \otimes {\bf 8_s} &= [0] \oplus [2] \oplus [4]_+
= {\bf 1} \oplus {\bf 28}  \oplus {\bf 35}_+ .
}}
Here $[n]$ denotes the $n$-times antisymmetrized representation of
$SO(8)$, with $[4]_+$ being self-dual.  Note that the representations
$[n]$ and $[8-n]$ are the same, being related by contraction with the
8-dimensional $\epsilon$-tensor.
The NS-NS and R-R spectra
together form the bosonic components of $D=10$ IIa (nonchiral) and IIb
(chiral) supergravity respectively.  In the NS-R and R-NS sectors are
the products
\eqn\prodrns{\eqalign{
{\bf 8_v} \otimes {\bf 8_c} &=
{\bf 8_s}\oplus{\bf 56_c} \cr
{\bf 8_v} \otimes {\bf 8_s} &=
{\bf 8_c}\oplus{\bf 56_s}.}
}
The $\bf 56_{s,c}$ are gravitinos, their vertex operators having one
vector and one spinor index.  They must couple to conserved spacetime
supercurrents.  In the IIa theory the two gravitinos (and
supercharges) have opposite chirality, and in the IIb the same.

Let us develop further the vertex operators for the R-R states.
This will involve a product of spin fields~\fms, $e^{-{{\varphi}\over{2}} -
{{\tilde{\varphi}}\over{2}} } S_{\alpha} \tilde{S}_{\beta}$.  These again
decompose into antisymmetric tensors, now of $SO(9,1)$:
\eqn\rrver{
V = e^{ -{{\varphi}\over{2}} - {{\tilde{\varphi}}\over{2}} } S_{\alpha}
\tilde{S}_{\beta}
( \Gamma^{[\mu_1} \cdots \Gamma^{\mu_n]}C)_{\alpha\beta} H_{[\mu_1 \cdots
\mu_n]}(X)
}
with $C$ the charge conjugation matrix.  In the IIa theory the
product is ${\bf 16} \otimes {\bf 16'}$ giving even $n$ (with $n \cong
10-n$) and in the IIb theory it is ${\bf 16} \otimes {\bf 16}$ giving odd
$n$.
As is usual, the classical equations of motion follow from the
physical state conditions, which at the massless level reduce to
$G_0 \cdot V = \tilde{G}_0 \cdot V = 0.$
The relevant part of $G_0$ is just $p_\mu \psi^\mu_0$ and similarly for
$\tilde G_0$.  The $p_\mu$ acts by differentiation on $H$, while
$\psi_0^\mu$ acts on the spin fields as it does on the corresponding
ground states: as multiplication by $\Gamma^\mu$.  Noting the identity
\eqn\gamma{
\Gamma^\nu \Gamma^{[\mu_1} \cdots \Gamma^{\mu_n]} =
\Gamma^{[\nu} \cdots \Gamma^{\mu_n]} +
\left( \delta^{\nu\mu_1} \Gamma^{[\mu_2} \cdots \Gamma^{\mu_n]}
+ {\rm perms} \right)}
and similarly for right multiplication, the physical state conditions
become
\eqn\bianchi{
dH=0 \qquad\qquad d{}^* H = 0.
}
These are the Bianchi identity and field equation for an antisymmetric
tensor field strength.  This is in accord with the representations found:
in the IIa theory we have odd-rank tensors of $SO(8)$ but even-rank
tensors of $SO(9,1)$ (and reversed in the IIb), the extra index being
contracted with the momentum to form the field strength.
It also follows that R-R amplitudes involving elementary strings vanish
at zero momentum, so strings do not carry R-R charges.

As an aside, when the dilaton background is nontrivial, the Ramond
generators have a term $\partial_\mu\phi \psi^\mu$, and the Bianchi
identity and field strength pick up terms proportional to
$d\phi \wedge H$ and $d\phi \wedge {}^*H$.  The Bianchi identity is
nonstandard, so $H$ is not of the form $dB$.  Defining $H' = e^{-\phi} H$
removes the extra term from both the Bianchi identity and field strength.
In terms of the action, the fields $H$ in the vertex operators appear
with the usual closed string $e^{-2\phi}$ but with non-standard dilaton
gradient terms.  The fields we are calling $H'$, which in fact are the
usual fields used in the literature, are decoupled from the dilaton.
This fact has played an important role in recent discussions of
string solitons and duality.

The IIb theory is invariant under world-sheet parity, so we can again
form an unoriented theory by gauging.
Projecting onto $\Omega =+1$ interchanges left-moving and right-moving
oscillators and so one linear combination of the R-NS and NS-R gravitinos
survives, leaving $D=10$, $N=1$ supergravity.  In the NS-NS sector, the
dilaton and graviton are symmetric under $\Omega$ and survive, while the
antisymmetric tensor is odd and is projected out. In the R-R sector, it
is clear by counting that the $\bf 1$ and ${\bf 35}_+$ are in the
symmetric product of ${\bf 8_s} \otimes {\bf 8_s}$ while the $\bf 28$ is
in the antisymmetric.  The R-R vertex operator is the product of right-
and left-moving fermions, so there is an extra minus in the exchange and
it is the $\bf 28$ that survives.  The bosonic massless sector is
thus ${\bf 1} \oplus {\bf 28}  \oplus {\bf 35}$, the $D=10$ $N=1$
supergravity multiplet.  This is the same multiplet as in the heterotic
string, but now the antisymmetric tensor is from the R-R sector.

The open superstring has only $N=1$ supersymmetry; in order that the
closed strings couple consistently they must also have $N=1$ supergravity
and so the theory must be unoriented.  In fact, spacetime anomaly
cancelation implies that the only consistent $N=1$ superstring is the
$SO(32)$ open plus closed string theory.  Now, as a general principle any
such inconsistency in the low energy should be related to some stringy
inconsistency.  This is the case, but it will be more convenient to
discuss this later after some discussion of $T$-duality.


\subsec{$T$-Duality of Type II Superstrings}
Even in the closed oriented Type II theories $T$-duality has an
interesting effect \refs{\dhs,\dbranes}.  Consider compactifying a single
coordinate $X^9$.  In the $R\to \infty$ limit the momenta are $p^9_R =
p^9_L$, while in the $R \to 0$ limit $p^9_R = -p^9_L$.  Both theories are
$SO(9,1)$ invariant but under {\it different} $SO(9,1)$'s.
Duality reverses the sign of the right-moving $X^9(\zb)$; therefore by
superconformal invariance it does so on $\tilde\psi^9(\zb)$.  Separate the
Lorentz generators into their left-and right-moving parts $M^{\mu\nu} +
\tilde M^{\mu\nu}$. Duality reverses all terms in $\tilde M^{\mu 9}$, so the
$\mu 9$ Lorentz generators of the $R\to 0$ limit are $M^{\mu 9} - \tilde
M^{\mu 9}$. In particular this reverses the sign of the helicity $\tilde
s_4$ and so switches the chirality on the right-moving side.  If one starts
in the IIa theory, with opposite chiralities, the $R\to 0$ theory has the
same chirality on both sides and is the IIb theory, and vice versa.  More
simply put, duality is a one-sided spacetime parity operation, and
reverses the relative chiralities of the right- and left-moving ground
states.  The same is true if one dualizes on any odd number of
dimensions, while dualizing on an even number returns the original Type
II theory.

Since the IIa and IIb theories have different R-R fields, $T_9$ duality
must transform one set into the other.  The action of duality on the spin
fields is of the form
\eqn\exa{\eqalign{
S_{\alpha} (z) &\to S_{\alpha} (z) \cr
 \tilde{S}_{\alpha} (\bar{z}) &\to \rho_9
   \tilde{S}_{\alpha} (\bar{z})
}}
for some matrix $\rho_9$.  In order for this to be consistent with the
action $\tilde\psi^9 \to -\tilde\psi^9$, $\rho_9$ must anticommute with
$\Gamma^9$ and commute with the remaining $\Gamma^\mu$.  Thus
$\rho =
\Gamma^9\Gamma^{11}$ (the phase of $\rho_9$ is determined, up to sign, by
hermiticity of the spin field).  Other
$\rho_m$ are similarly defined.  Now consider the effect on the R-R
vertex operators~\rrver.  The $\Gamma^{11}$ just contributes a sign,
because the spin fields have definite chirality.  Then by the
$\Gamma$-matrix identity~\gamma, the effect is to add a 9-index to $H$
if none is present, or to remove one if it is; the effect on the
potential $B$ ($H = dB$) is the same.  Take as an example the Type IIa
vector $B_\mu$. The component $B_9$ maps to the IIb scalar $B$, while
the $\mu\neq 9$ components map to $B_{\mu 9}$.  The remaining components
of $B_{\mu\nu}$ come from $B_{\mu \nu 9}$, and so on.

\subsec{$T$-Duality of Type I Superstrings}

The action of $T$-duality in the open and unoriented Type I theory
produces D-branes and orientifold planes, just as in the bosonic string.
Let us focus here on a single D-brane, taking a limit
in which the other D-branes and the orientifold planes are
distant and can be ignored.  Off the D-brane, only closed strings
propagate.  The local physics is that of the Type II theory, with two
gravitinos.  This is true even if though we began with the unoriented Type
I theory which has only a single gravitino.  The point is that the closed
string begins with two gravitinos, one with the spacetime
supersymmetry on the right-moving side of the world-sheet and one on the
left.  The orientation projection of the Type I theory leaves one linear
combination of these.  But in the $T$-dual theory, the
orientation projection does not constrain the local state of the string,
but relates it to the state of the (distant) image gravitino.
There are two independent gravitinos, with equal chiralities if an even
number of dimensions have been dualized and opposite if an odd number.

However, the open string boundary conditions are invariant under only one
supersymmetry.  In the original Type I theory, the left-moving world-sheet
current for spacetime supersymmetry $j_\alpha(z)$ flows into the boundary
and the right-moving current $\tilde j_\alpha(\bar z)$ flows out, so
only the total charge $Q_\alpha + \tilde Q_\alpha$ of the
left- and right-movers is conserved.  Under $T$-duality this becomes
$Q_\alpha + \prod_m \rho_m \tilde Q_\alpha$, the product running over all
the dualized dimensions.  Closed strings couple to open, so the general
amplitude has only one linearly realized supersymmetry.  That is, the
vacuum without D-branes is invariant under $N=2$ supersymmetry, but the
state containing the D-brane is invariant under only $N=1$: {\it it is a
BPS state \joetwo.}
\nref\blackp{G. T. Horowitz and A. Strominger,
Nucl. Phys. {\bf B360} (1991) 197.}
\nref\hullt{C. M. Hull and P.K. Townsend,
Nucl. Phys. {\bf B438} (1995) 109.}
\nref\wit{E. Witten, Nucl. Phys. {\bf B443} (1995) 85.}
\nref\coni{A. Strominger, Nucl. Phys. {\bf B451} (1995) 96.}

BPS states must carry conserved charges.  In the present case there is
only one set of charges with the correct Lorentz properties, namely the
antisymmetric R-R charges.  The world volume of a $p$-brane naturally
couples to a ($p$$+$$1$)-form potential $A_{p+1}$, which has a
($p$$+$$2$)-form field strength $F_{p+2}$.  This identification can
also be made from the $g^{-1}$ behavior of the D-brane tension: this is the
behavior of an R-R soliton~\refs{\blackp,\hullt,\wit,\coni}.
\nref\joeandy{
J. Polchinski and A. Strominger, {\it New Vacua for Type II String Theory,}
preprint UCSBTH-95-30, NSF-ITP-95-136, hep-th/9510227 (1995).}

The IIa theory has $p = 0$, 2, 4, 6, and 8-branes.  The vertex
operators~\rrver\ describe field strengths of all even ranks.  By a
$\Gamma$-matrix identity the $n$-form and $(10-n)$-form field strengths
are Hodge dual to one another, so a $p$-brane and $(6-p)$-brane are
sources for the same field, but one magnetic and one electric.  The field
equation for the 10-form field strength allows no propagating states, but
the field can still have a physically significant energy
density~\refs{\joetwo,\joeandy}.  Curiously, the 0-form field strength should
couple to a $(-2)$-brane, but it is not clear how to interpret this.

The IIb theory has $p = -1$, 1, 3, 5, 7, and 9-branes.  The vertex
operators~\rrver\ describe field strengths of all odd ranks, appropriate
to couple to all but the 9-brane.  A $(-1)$-brane is a Dirichlet instanton,
defined by Dirichlet conditions in the time direction as well as all
spatial directions.  The 9-brane does couple to a nontrivial
{\it potential,} as we will see below.

The action
for a ($p$$+$$1$)-form potential takes the form
\eqn\formaction{
S={1\over2}\int F_{p+2}{}^*F_{p+2}+i\mu_p\int_{p-\rm branes} A_{p+1},
}
where the ($p$$+$$1$)-form charge is $\mu_p$.\foot
{This is not correct for $p=3$, for which the field strength is self-dual.
There is no covariant action in this case.}
In addition the coupling of the D-brane to NS-NS and open string states
has the same form~\branetheory\ as the bosonic D-brane theory.

It is interesting to consider the effect of
$T$-duality.  Consider a $p$-brane, which couples to the R-R potential
with $p+1$ indices tangent to the brane world-sheet.  Take the $T$-dual
in a direction $\mu$ perpendicular to the D-brane.  The Dirichlet
condition in this direction becomes Neumann, so in the dual theory there
is a $(p+1)$-brane.  At the same time, as discussed at the end of
section~3.3, the R-R potential acquires an extra $\mu$ index, as needed
to couple to the $(p+1)$-brane.  Similarly, if we take the $T$-dual in a
direction $m$ along the brane, it becomes a $(p-1)$-brane and the R-R
potential loses its $m$ index.

The D-brane, unlike the fundamental string, carries R-R charge.  It
is interesting to see how this is consistent with our earlier discussion
of string vertex operators (this argument was first given by Bianchi,
Pradisi, and Sagnotti~\ref\bpsag
{M. Bianchi, G. Pradisi, and A. Sagnotti, Nucl. Phys. {\bf B376} (1992)
365.}). The R-R vertex operator~\rrver\ is in the
$(-\ha ,-\ha)$ picture, which can be used in almost all processes.  In
the disk, however, the total right+left ghost number must be $-2$.  With
two or more R-R vertex operators, all can be in the $(-\ha ,-\ha)$
picture (with picture changing operators included as well), but a single
vertex operator must be in either the $(-{3\over 2}, -\ha)$ or the
$(-\ha,-{3\over 2})$ picture.  The  $(-\ha ,-\ha)$ vertex operator is
essentially $e^{-\varphi} G_0$ times the $(-{3\over 2}, -\ha)$ operator,
so besides the shift in the ghost number the latter has one less power of
momentum and one less $\Gamma$-matrix. The missing factor of momentum
turns $H$ into $A$, and the missing $\Gamma$-matrix gives the correct
Lorentz representations for the potential rather than the field strength.
\nref\doug{M. R. Douglas, {\it Branes
within Branes,} preprint RU-95-92, hep-th/9512077.}
\nref\csapp{M. Bershadsky, C. Vafa, and V. Sadov, {\it D-Branes and
Topological Field Theories,} preprint HUTP-95-A047,
hep-th/9511222;\hfil\break  A. Strominger, {\it Open $p$-Branes,}
preprint hep-th/9512059.}

For open string gauge fields, the $-1$ picture involves the potential and
the $0$ picture the field strength.
\nref\li{
M. Li, {\it Boundary States of D-Branes and Dy-Branes,} preprint
BROWN-HET-1020, hep-th/9510161.}In interactions
involving one R-R field and $k$ open string gauge fields, in order for
the pictures to add to $-2$, exactly one vector or else the R-R field
must appear as a potential, and the rest as field strengths.
Thus these interactions are of Chern-Simons form~\rrex.  Their detailed
form has been discussed recently~\refs{\li,\doug}.  All interactions
\eqn\csint{
\int A_k F^l }
having the correct rank to be integrated over the $p$-brane world-sheet
(that is, $k + 2l = p+1$) appear.
These have
played an important role in various recent discussions of D-brane
dynamics~\refs{\witbound,\bound,\csapp}.

To obtain the D-brane tension and R-R charge, one can consider the same
vacuum cylinder as in the bosonic string~\rrex.
Carrying out the traces over the open superstring spectrum
gives
\eqn\spotential{\eqalign{
A = & 2V_{p+1}  \int {dt\over 2t}\, (8\pi^2 \ap t)^{-(p+1)/2}
 e^{- t{Y^2\over2\pi \alpha'}} \prod_{n=1}^\infty (1-q^{2n})^{-8}\cr
  & \qquad\qquad {1\over2} \left\{ -f_2(q)^{8}
   + f_3(q)^8 - f_4(q)^8 \right\},}}
where again $q =e^{-\pi t}$.
The three terms in the braces come from the open string R
sector with ${1\over2}$ in the trace, from the NS sector with
${1\over2}$ in the trace, and the NS sector with ${1\over2} (-1)^F$
in the trace; the R sector with ${1\over2} (-1)^F$ gives no net
contribution.  These three terms sum to zero by the `abstruse
identity,' because the open string spectrum is supersymmetric.
In terms of the closed string exchange, this
reflects the fact that D-branes are BPS states, the net forces from
NS-NS and R-R exchanges canceling.  The separate exchanges can be
identified as follows.  In the terms with $(-1)^F$, the world-sheet
fermions are periodic around the cylinder corresponding to R-R
exchange, while the terms without $(-1)^F$ have antiperiodic fermions and
come from NS-NS exchange.  Obtaining the $t\to 0$ behavior as before
gives
\eqn\limit{\eqalign{
{\cal A} \sim & {1\over2} (1-1) V_{p+1}  \int{dt\over t} (2\pi
t)^{-(p+1)/2} (t/2\pi\alpha')^4 e^{- t{Y^2\over8\pi^2 \alpha'^2}} \cr
= & (1-1) V_{p+1} 2\pi (4\pi^2\alpha')^{3-p} G_{9-p}(Y^2).
}}
The $(1-1)$ is from the NS-NS and R-R exchanges respectively.
Comparing with field theory calculations gives~\joetwo
\eqn\look{
\mu_p^2 = 2 T_p^2 = 2\pi (4\pi^2\alpha')^{3-p}.
}

Just as a constant Wilson line is dual to a translation of the D-brane, a
constant field strength is dual to a rotation or boost \ref\bachas{C.
Bachas, {\it D-Brane Dynamics,} preprint
NSF-ITP-95-144, hep-th/9511043.}.  D-branes which are not parallel feel a
net force because the cancelation is no longer exact.  In the extreme case,
where one of the D-branes is rotated by $\pi$, the coupling to the dilaton
and graviton is unchanged but the coupling to the R-R tensor is reversed in
sign, and the two terms in the cylinder amplitude add.  In fact, a
well-known divergence of Dirichlet boundary conditions sets in for
non-parallel branes: the
$t$-integration diverges at zero.  This is similar to the Hagedorn
divergence, and represents an instability of the D-branes when brought too
close~\ref\tomlen{T. Banks and L. Susskind, {\it Brane - Anti-Brane Forces,}
preprint RU-95-87, hep-th/9511194.}.

The orientifold planes also break half the supersymmetry and are R-R and
NS-NS sources.  In the original Type I theory the orientation projection
keeps only the linear combination $Q_\alpha + \tilde Q_\alpha$; in the
dualized theory this becomes $Q_\alpha + \prod_m \rho_m
\tilde Q_\alpha$ just as for the D-branes.  The force between an
orientifold plane and a D-brane can be obtained from the M\"obius strip as
in the bosonic case; again the total is zero and can be separated into
NS-NS and R-R exchanges.  The result is similar to the bosonic
result~\orten, \eqn\ormu{ \mu'_p = \mp 2^{p - 5} \mu_p, \qquad T'_p = \mp
2^{p - 5} T_p }
Since there are $2^{9-p}$ orientifold planes, the total fixed-plane charge
is $\mp 16 \mu_p$, and the total fixed-plane tension is $\mp 16 T_p$.

A nonzero total tension represents a source for the graviton and dilaton,
so that at order $g$ these fields become time dependent as in the
Fischler-Susskind mechanism~\ref\fsuss{W. Fischler and L. Susskind, Phys.
Lett. {\bf B171} (1986) 383; {\bf 173} (1986) 262.}.
A nonzero total R-R source is more serious: the field equations are
inconsistent, because R-R flux lines have no place to go in the compact
space.\foot
{The Chern-Simons coupling~\csint implies that the open
string field strengths are also R-R sources, so there will be more
general consistent solutions with nonzero values for these.}  So we need
exactly 16 D-branes with the $SO$ projection, giving the $T$-dual of
$SO(32)$.  So we find that the spacetime anomalies for $G \neq SO(32)$
are accompanied by a divergence~\ref\gsdiv{M. B. Green and J. H. Schwarz,
Phys. Lett. {\bf B149} (1984) 117;\hfil\break {\bf 151B} (1985) 21.};
this also leads to a world-sheet conformal anomaly that cannot be
canceled because of the inconsistency of the field equations.  All this
can be discussed in the original $D=10$ Type I theory~\rrex.  The Neumann
open strings correspond to 9-branes, since the endpoints can be
anywhere.  The Dirichlet and orientifold 9-branes couple to an R-R
10-form,
\eqn\tenform{ i (32 \mp N) {\mu_{10}\over 2} \int A_{10} ,}
and the field equation from varying~$A_{10}$ is $G = SO(32)$~\rrex!

\newsec{Lecture IV:  {\sl D-Branes Galore}}

\subsec{Discussion}

We have seen that $T$-duality of the Type I string leads to a theory with
precisely 16 Dirichlet $p$-branes on a $T_{9-p}/Z_2$ orientifold, for any
given value of $p$.
We now understand that the restriction to 16 comes
from conservation of R-R charge.  It follows that in a non-compact space,
where the flux lines could run to infinity, we could have a consistent
theory with any number and configuration of $p$-branes, with all $p$
being even in the IIa theory or odd in the IIb.  Indeed, cluster
decomposition plus $T$-duality forces this upon us.  The $T$-dual of a
flat torus gives flat D-branes, but because they are dynamical this is
continuously connected to configurations where the D-branes fold back and
forth, and in this way one can reach a configuration which in any local
region has an arbitrary set of $p$-branes.  Moreover, while $T$-duality
gives at first only $p$-branes for a single value of $p$, we can then
deform to a configuration with perpendicular $p$-branes.  A further
$T$-duality along a direction which is parallel to one $p$-brane and
perpendicular to another
interchanges Neumann and Dirichlet conditions along that
direction, and so produces a $(p$$+$$1)$-brane and a $(p$$-$$1)$-brane.
In this way we reach a general configuration.

Thus it is natural to consider all these configurations as different
states in a single theory, with the usual Type I and II strings being
perturbative expansions around particular states (the latter being the
no-brane state).  There is an important consistency check here.  The
field strengths to which a $p$-brane and $(6$$-$$p)$-brane couple are
dual to one another, $H_{p+2} = {}^*H_{10-p}$.  This implies a Dirac
quantization condition, as generalized by Teitelboim and
Nepomechie~\ref\dirac{R. I. Nepomechie, Phys. Rev. {\bf D31} (1985)
1921;\hfil\break C. Teitelboim, Phys. Lett. {\bf B167} (1986) 63, 69.}.
Integrating the field strength ${}^*H_{p+2}$ on an ($8$$-$$p$)-sphere
surrounding a $p$-brane, the action \formaction\ gives a total flux $\Phi=
\mu_p$.  We can write
${}^*H_{p+2} = H_{8-p} = d B_{7-p}$ everywhere except on a Dirac `string'.
Then  \eqn\then{
\Phi= \int_{S_{8-p}} {}^*F_{p+2}=\int_{S_{7-p}} B_{7-p},
}
where we perform the last integral on a small sphere surrounding the Dirac
string.  A ($6$$-$$p$)-brane passing circling the string picks up a phase
$e^{i \mu_{6 - p}\Phi}$.  The condition that the string be invisible is
\eqn\dquant{
\mu_{6 - p} \Phi = \mu_{6 - p} \mu_p = 2\pi n.}
The D-branes charges~\look\ satisfy this with the minimum quantum
$n=1$.\foot
{This argument does not apply directly to the case $p=3$, as the
self-dual 5-form field strength has no covariant action.  However, using
$T$-duality to relate this to $p=2$ shows that the $p=3$ quantum is
minimal also.}

This calculation has the look of a `string miracle.'  It is not at all
obvious why the one-loop open string calculation should have given just
this result.  Had the R-R charges not satisfied the quantization
condition, one could likely use the argument from the first paragraph of
this section to show that the Type I theory has some sort of
non-perturbative anomaly.  Perhaps this can be used to find a more direct
topological calculation of the D-brane charge.
\nref\rrquanta{J. A. Harvey and A. Strominger, Nucl. Phys. {\bf B449}
(1995) 535.}\nref\rrquantb{C. Vafa and E. Witten, Nucl. Phys. {\bf B447}
(1995) 261.}

Thus far weak/strong coupling string duality has not entered.  All we have
done is to follow $T$-duality to its logical conclusions,
though not all of these were noticed until string duality focused
attention on the important issues.  Now, the key point is that string
duality relates ordinary strings (which carry electric NS-NS charges),
as well as string solitons carrying magnetic NS-NS charges, to R-R
charged states~\refs{\hullt,\wit}.  The D-brane description of the R-R
charged states has allowed many new and successful tests of string
duality.  Most fundamentally, string duality makes a specific prediction
for the quantum of R-R charge~\refs{\rrquanta,\rrquantb}, which is precisely
the value~\look\ carried by the D-brane~\joetwo.\foot
{To be precise, there remains a factor of two discrepancy in the
literature, which can plausibly be attributed to the problem of defining
the action for the chiral bosons of the string soliton~\rrquanta.}
In the remainder of this lecture a few additional consequences will be
derived.

Before the observation that D-branes carry R-R charge, the R-R charged
states required by string duality were assumed to be black holes.  One
can always find such black hole solutions~\blackp.  What is the
relation between these descriptions?  My understanding is as follows.
Because the dilaton scales out of the R-R action, the R-R solitons are
small, their size being given by the Planck scale~\ref\shenkscale{S. H.
Shenker, {\it Another Length Scale in String Theory?} preprint RU-95-53,
hep-th/9509132 and seminar at ITP, Jan. 1996.}.  For weak string coupling
this is smaller than the string length, so the nonlinear part of the black
hole solution is just not relevant, and the D-brane is actually a small
perturbation on the geometry.  This is consistent with the discussion of
single-fermion tunneling in the matrix model~\ref\shenkemg{S. H. Shenker, in
{\it Cargese 1990, Proceedings:  Random Surfaces and Quantum Gravity} (1990)
191.}, which is also an effect of order $g^{-1}$, and so is a small
disturbance as compared to a normal field theory tunneling event.  A rather
opposite interpretation\foot{Suggested by E. Witten and A. Strominger.} is
that an open string ending on a D-brane is actually a closed string, half
of which is stuck behind the horizon of a black hole!  This is curiously
similar to the picture of the black hole entropy in
ref.~\ref\sussug{L.~Susskind and J.~Uglum, Phys. Rev. {\bf D50} (1994)
2700.}.

Taking this issue further, for sufficiently many coincident D-branes
string perturbation theory will break down, the expansion parameter being
$gN$.  However, for large enough R-R charge, the description in terms of
low energy field theory becomes valid because the black hole is large.
In some cases it is possible to continue between these regimes by varying
parameters, and to follow the BPS states.\foot{
The BPS stategy was applied to Neveu-Schwarz black holes by Larsen and
Wilczek~\ref\larwil{F. Larsen and F. Wilczek, Phys. Lett. {\bf B375}
(1996) 37.}.  In this case the black hole continue to look like a black
hole no matter how weak the coupling becomes, and so one does not have
an explicit understanding of the space of states even at weak coupling.}
Very recently, this has
led to the counting of the BPS states of a black
hole~\ref\stromvafa{A. Strominger and C. Vafa, {\it Microscopic Origin of
the Bekenstein-Hawking Entropy,} preprint HUTP-96-A002,
hep-th/9601029.}, and the number is indeed that given by the
Bekenstein-Hawking entropy.  This is the first time, after two decades of
attempts, that the black hole entropy has been related to a counting of
states in a controlled way.

\subsec{Multiple Branes and Broken Supersymmetries}

$T$-duality of the Type I string lead to parallel D-branes with given $p$.
This configuration has the same supersymmetry as the original Type~I theory.
For convenience we will in this section use $D=4$ units, so this is $N=4$
SUSY, broken from the $N=8$ of the Type~II theory.

Now we are considering more general configurations of D-branes, and so will
determine the unbroken supersymmetry of such configurations (there is some
discussion of this in ref.~\ref\bsv{M. Bershadsky, C. Vafa, and V. Sadov,
{\it D-Strings on D-Manifolds,} preprint HUTP-95-A035, hep-th/9510225.}).
For simplicity we will analyze only the case that all D-branes are oriented
along some set of coordinate axes, so each can be defined by taking
Dirichlet boundary conditions on some subset $S_i\subset\{0,1,\ldots,9\}$ of
the coordinates ($i$ labeling the D-brane) and Neumann conditions on the
remaining coordinates $\overline S_i$.  We can use $T$-duality to simplify
the discussion.  By dualizing on each of the axes in $S_1$ we can take the
first D-brane to be a nine-brane, with fully Neumann conditions.  Now
consider a second D-brane.  From earlier discussion, we know that the
supersymmetries left unbroken by the two D-branes are respectively
\eqn\unbroken{
Q_\alpha+ {\tilde Q}_\alpha, \qquad
Q_\alpha+ \rho_{(2) \alpha\beta}
  {\tilde Q}_\beta\ .
}
The unbroken supersymmetries are the intersections of these two sets, and
therefore are in one-to-one correspondence with the $+1$ eigenvalues of
$\rho_{(2)}=\prod_{m\in S_2} \left(\Gamma^m\Gamma^{11}\right)$.
Note that $S_2$ has an even number of elements,
because we must now be in the IIb theory.  For the case of two elements,
meaning that the second D-brane is a seven-brane, there are no unbroken
supersymmetries: we have $(\Gamma^m\Gamma^{11}\Gamma^{m'}\Gamma^{11})^2
= -1$ so the eigenvalues of $\rho_{(2)}$ are $\pm i$.  For four elements,
$\rho_{(2)}^2 = +1$; half the eigenvalues are $+1$ and so the unbroken
supersymmetry is $N=2$.  Similarly, six elements break all the
supersymmetry, and eight break half.

We can state the above result in a $T$-duality invariant way.  Consider open
strings with one end on one D-brane and one on the other.  Some coordinates
will have Neumann conditions on both ends (NN), some Dirichlet (DD), and
some mixed (ND).  Unbroken supersymmetry requires the number of ND
directions to be a multiple of 4.  We can also see this as follows.  The
open string mode expansion is
\eqn\modep{
X(0,\sigma_1) \sim i {\sqrt{{\alpha'}\over{2}}}
 \sum_r {{\alpha_r}\over{r}} ( e^{i r \sigma_1}
   \pm e^{-ir \sigma_1} ).
}
Here $r$ is integer for NN and DD coordinates, with the upper and lower sign
respectively (the $r=0$ terms have not been written explicitly), while $r$
is half-integer for ND and DN coordinates.  The fermions have the same
moding in the R sector and opposite in the NS sector.  Let $\nu$ be the
number of ND coordinates.  The string zero point energy is 0 in the R sector
as always, and
\eqn\ndzpe{
(8-\nu)\left(-{1\over 24} - {1\over 48}\right)
+ \nu \left({1\over 24} + {1\over 48}\right) = -{1\over 2} + {\nu\over 8}
}
in the NS sector.
Only for $\nu$ a multiple of 4 is degeneracy between the R and NS sectors
possible.

Similarly, one can show that a nine-brane plus two five-branes, with $S_2=
\{6,7,8,9\}$ and $S_3 = \{4,5,8,9\}$, break the supersymmetry down to
$N=1$. This is $T$-dual to three five-branes (with $S_1=\{4,5,6,7\}$,
$S_2=
\{4,5,8,9\}$ and $S_3 = \{6,7,8,9\}$); also to three seven-branes, and so
on.

The story above cannot be complete.\foot
{The remainder of this section is a result of discussions between J.
Harvey, G. Moore, J. Polchinski, and A. Strominger.}
A $p$-brane and
$(p+2)$-brane, when separated, do indeed break all supersymmetries.
However string duality requires that they have a bound state where the
$p$-brane is fully contained in the $(p+2)$-brane, with as much
supersymmetry as the
$(p+2)$-brane has by itself.  We can see a hint of this by considering
open strings with one end on each D-brane.  The zero point energy found
above is negative for these.  When the $p$- and
$(p+2)$-branes are well separated, the energy of stretching makes the
open strings massive, but for sufficiently small separation there are
tachyonic open strings, not all of which are removed by the GSO
projection.  This is like the Hagedorn instability, and cannot be
quantitatively treated.  However, in this case one can see in
another description that there is a stable BPS state to decay to.  The
$(p+2)$-brane has a world-sheet gauge field.  Consider a constant
background for its field-strength $F_{mn}$.  The
Chern-Simons coupling~\csint,
\eqn\flux{
\int A_{p+1} F
}
implies that the $(p+2)$-brane now couples to the $(p+1)$-form potential
as well as the usual $(p+3)$-form.  That is, it has the total R-R quantum
numbers of the $(p+2)$-brane and the $p$-brane.  In effect the $p$-brane
has dissolved in the $(p+2)$-brane!

This discussion is exactly parallel
to the discussion of the binding of fundamental and D-strings in
ref.~\witbound.  In fact it is dual to it.  By $T$-duality we may
consider a D three-brane and D one-brane in Type IIB theory.  This theory
has an $SL(2,Z)$ self-duality which takes the three-brane into itself
and the D one-brane into a fundamental string.  Now $T$-dualize in the two
directions parallel to the three-brane and perpendicular to the string.
The result is parallel Dirichlet and fundamental strings, the case
considered in ref.~\witbound.

\subsec{D-strings}

The only supersymmetric objects in the Type~I theory, besides the
nine-branes, will be one-branes and five-branes.
This is consistent
with the fact that the only R-R field strengths are the three-form and its
Hodge-dual seven-form.\foot
{The binding of $p$ and $p+2$ is therefore not relevant here.}  To close
these lectures we will study these objects in the Type IIb and Type I
theories.

Start with the Dirichlet one-brane, or
D-string, first in the Type IIb theory and then in the Type I theory.
In the Type II string, a $p$-brane is just the $T$-dual
of the Type I nine-brane.  In particular, its massless sector is just the
dimensional reduction of the $D=10$, $N=1$ gauge multiplet to $p+1$
dimensions.  Thus we have the bosonic states
\eqn\photon{
\psi_{-1/2}^\mu |{ k}, ij\rangle,\ \mu=0,\ldots,p \qquad \psi_{-1/2}^m
|{ k}, ij\rangle,\ m=p+1,\ldots,9\
}
where $k^{\mu}$ is a $(p+1)$-dimensional momentum vector.  The
spacetime spinor of $SO(8)$ is projected along Dirichlet and Neumann
directions, under an $SO(9-p)$$\times$$SO(p+1)$ decomposition.

For the Type IIb D-string, $p=1$, the gauge field has no local dynamics,
so the only bosonic excitations are the transverse fluctuations.  Applying
the GSO projection (e.g.~via locality with the gravitino vertex operator),
the right-moving spinors on the D-string are in the $\bf 8_s$ of $SO(8)$,
and the left-moving spinors in the $\bf 8_c$.  This is the same as the
world-sheet theory of a macroscopic fundamental IIb string \witbound!  This
is as required by weak/strong self-duality of the IIb string
\refs{\hullt,\wit}.  The fundamental and D-strings couple respectively to
NS-NS and R-R  two-form potentials, which are interchanged by weak/strong
duality.  Their tensions are respectively $O(1)$ and $O(e^{-\phi})$ in the
string metric, which become $O(e^{\phi/2})$ and $O(e^{-\phi/2})$ in the
Einstein metric, and so are interchanged under $\phi \to -\phi$.  Indeed,
given the argument that D-branes must appear in the Type II spectrum, the
BPS bound implies that at strong coupling they are the lightest degrees of
freedom.  This strongly suggests that the physics in this limit is given by
an effective theory of D-strings: string duality.\foot{It does not imply
that duality holds to all energies, but this is the simplest possibility.
That is, given that physics below the Planck energy is described by some
specific string theory, it seems likely that there is a unique extension to
higher energies.}

The full duality group of the $D=10$ Type IIb theory is believed to be
$SL(2,Z)$~\refs{\hullt,\wit}.  This relates the fundamental string not only
to the R-R string but to a whole set of strings $(m,n)$ for $m$ and $n$
relatively prime~\ref\iibstring{J. H. Schwarz, Phys. Lett.
{\bf B360} (1995) 13; (E) {\bf B364} (1995) 252.}.  Here $m$ and $n$ are
respectively the NS-NS and R-R tensor charges of the string.  Witten has
used the D-brane picture to show that these strings exist as bound states of
$n$ D-branes and
$m$ fundamental strings~\witbound.  The non-dynamical $U(n)$ gauge
symmetry of $n$ coincident one-branes plays an essential role here.  This
method has been generalized successfully to a number of other counting
problems~\refs{\witbound,\bound}, most notably the black hole
entropy~\stromvafa\ discussed above.

Now let us move on to the one-brane of the Type I theory \ref\polwit{
J. Polchinski and E. Witten {\it Evidence for Heterotic - Type I Duality,}
preprint IASSNS-HEP-95-81, hep-th/9510169.}.  There are two modifications.
The first is the projection onto oriented states.  The $U(1)$ gauge field,
with vertex operator $\partial_t X^\mu$, is removed just as the vector of the
Type~I string spectrum.  The collective coordinates, with vertex operators
$\partial_n X^\mu$, remain in the spectrum because the normal derivative is
even under $\Omega$.  That is, in terms of its action on the $X$ oscillators
$\Omega$ has an additional $-1$ for the $m=2,\ldots,9$
directions, as compared to the action on the usual NN strings.  By
superconformal symmetry this must extend to the fermions, so that on the
ground states $\Omega$ is no longer the identity but acts as
$R=e^{i\pi(S_1+S_2+S_3+S_4)}$.  This removes the left-moving $\bf 8_c$ and
leaves the right-moving $\bf 8_s$ (or vice versa: we have made an arbitrary
choice in defining $R$).

The second modification is the inclusion of 1-9 strings, strings with
one end on the one-brane and one on a nine-brane, the latter corresponding
to the usual $SO(32)$ Chan-Paton factor.

\vskip 0.5cm
\hskip 2.4cm\epsfxsize3.3in\epsfbox{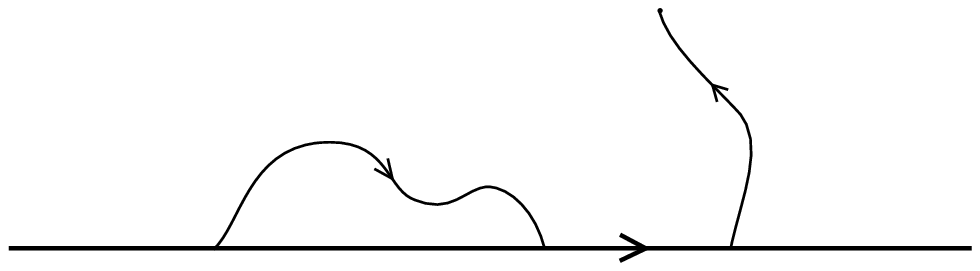}
\vskip 0.5cm

The $\Omega$
projection determines the 9-1 state in terms of the 1-9, but
otherwise makes no constraint.  The calculation~\ndzpe\ shows that
there are no massless states in the NS sector.  The R ground states are, as
always, massless.  Here there are two, from the periodic $\psi^{0,1}$
oscillators,
\eqn\ram{
|\pm; i\rangle = ( \psi^0_0 \pm \psi^1_0 ) |i\rangle,
}
where $i$ is a Chan-Paton index for the nine-brane end. One of the
two states
$|\pm; i\rangle$ is
removed by the GSO projection, and the $G_0$ physical state condition
then
implies that these massless fermions are chiral on the one-brane.
Spacetime
supersymmetry can only be satisfied if the GSO projection is such
that they move oppositely to the 1-1 fermions; at the world-sheet level
this would have to follow from a careful analysis of the OPE of the
gravitino. The 1-9 strings, with one Chan-Paton index, are vectors of
$SO(32)$.

Thus the world-sheet theory of the Type I one-brane is precisely that of
the heterotic string,
with the spacetime supersymmetry realized in Green-Schwarz form and the
current algebra in fermionic form.  This is again strong evidence for
string duality, here between the $SO(32)$ Type I and heterotic strings.
Curiously, the same conclusion follows from the black one-brane
description~\ref\blackone{A. Dabholkar,
Phys. Lett. {\bf B357} (1995) 307;\hfil\break C. M. Hull,
Phys. Lett. {\bf B357} (1995)
545.}, even though the details are quite different.

The fermionic $SO(32)$ current algebra requires a GSO projection.  It is
interesting to see how this arises in the D-string.  Consider a closed
D-string.  The $\Omega$ projection removed the $U(1)$ gauge field, but is
consistent with a discrete gauge symmetry, a holonomy $\pm 1$ around the
D-brane.  This discrete gauge symmetry is the GSO projection, and evidently
the rules of D-branes require us to sum over all consistent possibilities in
this way.

We can now see how D-strings account for the spinor
representation of $SO(32)$ in the Type~I theory.  In the {R} sector of
the discrete D-brane gauge theory, the
1-9 strings are periodic.  The zero modes of the fields $\Psi^i$,
representing the massless
1-9 strings, satisfy the Clifford algebra
\eqn\oneb{ \{ \Psi^i_0 , \Psi^j_0 \} = \delta^{ij}, \qquad i,j= 1,
\cdots , 16 .}
The quantization now proceeds just as for the fundamental heterotic string,
giving spinors ${\bf 2^{15}} + \overline{\bf  2^{15}}$.

We can follow this further, looking for the $E(8) \times E(8)$ Type I
string (much of the following is based on ref.~\polwit\ and discussions
with E. Witten).  Let us start with a single $E(8)$.  Compactify the $SO(32)$
heterotic string on a circle, with $U(1)^{16} \subset SO(32)$ Wilson line
\eqn\ewilson{ \left({\ha}^7,0^9\right).
}
This breaks the $SO(32)$ to $SO(14) \times
SO(18)$.  As the radius is reduced, massless winding states appear at $R^2 =
{1\over 8} {\ap}$.  Winding numbers $\pm 1$ contribute spinors
$({\bf 64},+1)$ and $(\overline{\bf 64},-1)$ of $SO(14) \times U(1)$, the
$U(1)$ being the left-moving Kaluza-Klein momentum.  Winding numbers $\pm
2$ contribute $({\bf 14},\pm 2)$.  These add up to the adjoint of $E(8)$.

In the Type I theory the winding states map to D one-branes, and we should
be able to find all these states.  But there is a paradox~\polwit.
In the Type I theory, the $E(8)/SO(14)\times U(1)$ gauge bosons are
D-branes, not perturbative string states.  With all the recent work on
supersymmetric gauge theories and string theories, we have gotten used to
something that once seemed unlikely: nonperturbative states can become
massless at special values of the parameters.  But in all known examples,
this happens {\it only when perturbation theory breaks down}.  This is
consistent with the idea that perturbation theory is in some sense
asymptotic (at fixed energy)---light nonperturbative states would
violate this.  In field theory one can probably prove it.  In string
theory we have no nonperturbative formulation on which to base a proof, but
it seems likely and is consistent with all examples.  Now apply this to the
present case.  Hold the heterotic string radius fixed at the symmetry point,
$R_{\rm h}= \sqrt{\ap/8}$ and take the coupling $g_{\rm h}$ large.
According to the Type~I--heterotic duality, the corresponding Type I theory
has
\eqn\onehet{
R_{\rm I} = R_{\rm h} g_{\rm h}^{-1/2}, \qquad g_{\rm I} = g_{\rm h}^{-1}\ .
}
In the limit of interest $R_{\rm I}$ is becoming small, so the physics will
be clearer in the $T$-dual theory, which we will denote I$'$.  Then,
taking into account the transformation~\dilt\ of the dilaton,
\eqn\onephet{
R_{\rm I'} = R^{-1}_{\rm h} g_{\rm h}^{1/2}, \qquad g_{\rm I'} = g_{\rm
h}^{-1/2} R^{-1}_{\rm h}\ . }
The limit $g_{\rm h} \to \infty$, $R_{\rm h}$ fixed then corresponds to a
weakly coupled Type I$'$ theory on a large $S_1/Z_2$, where one would not
expect to see a massless soliton.

The resolution involves the special features of the Type I$'$ theory.
We know that the Wilson line~\ewilson\ puts seven D-branes at one fixed plane
and nine at the other.  The R-R charge of the fixed planes is canceled
globally but not locally (the latter, by symmetry, would require eight at
each end).  So the spacetime is an R-R capacitor, with a net source at one
end and net sink at the other.  But by the BPS property there are also local
dilaton sources, so there is a dilaton gradient.  This gradient is of order
$g_{\rm I'}^{1}$, but $R_{\rm I'}$ is of order $g_{\rm I'}^{-1}$  so the
effect is of order~1.  The precise dilaton dependence is obtained by solving
the field equations of the effective supergravity theory.  This is the IIa
supergravity theory, since we have dualized one dimension.  The R-R
background is a nine-form potential, which is non-dynamical in $D=10$ but
contributes an effective cosmological constant.  This is the supergravity
theory found by Romans~\ref\romans{L. J. Romans, Phys. Lett. {\bf B169}
(1986), 374.}.  The solution was found in ref.~\polwit.  It  has the
following property.  As $R_{\rm h}$ is decreased toward the $E(8)$ radius,
$R_{\rm I'}$ increases and so does the effect of the dilaton gradient.
Precisely at the critical radius, the dilaton diverges at the end with seven
D-branes.  This is so even though the effective nine-dimensional coupling,
involving some average of the dilaton, remains weak.

The paradox is thus evaded, and the precise point of breakdown gives
further evidence for Type~I--heterotic duality.  We can go further and find
the $E(8)/SO(14)\times U(1)$ gauge bosons in the Type I$'$ D-brane
spectrum.  In the heterotic theory these are winding states, so one-branes
in the Type I theory and zero-branes in the Type I$'$.  The winding number
one states map into single zero-branes, which by symmetry must be at one
fixed plane.  From the relation~\wherends,
\eqn\ends{
X'^9(\pi,\sigma^2) - X'^9(0,\sigma^2) = 2\pi\ap p^{9},
}
one can deduce that the one-branes from the current algebra R sector map to
zero-branes at the end with nine eight-branes, and those from the current
algebra NS sector to the end with seven.  It is the latter that are of
interest.  These have a mass of order $e^{-\phi}$, which does indeed go to
zero at this end when the radius becomes critical.  Of course, we cannot
follow the state all the way to strong coupling, but in the range where
the coupling is still weak this is a BPS state and the supergravity
solution for $\phi$ gives the mass required by duality.  The reader can work
out the one-brane spectrum just as done for the zero-brane above, and it is
as expected.  It is free to move on the fixed plane but not away from it, and
the 0-8 strings give rise to a spinor representation of
$SO(14)$.  The $U(1)$ is the R-R gauge field which couples to zero-branes.
The remaining gauge bosons have winding number two and so map to a pair of
zero-branes.  The necessary bound states can be studied as in
refs.~\refs{\witbound,\bound}, but we have not carried this out in detail.

For the $E(8) \times E(8)$ theory the Wilson line is~\ref\gins{P. Ginsparg,
Phys. Rev. {\bf D35} (1987) 648. }
\eqn\eewilson{
\left({\ha}^7, \ha-\lambda, \lambda, 0^7 \right)
}
with critical radius $R_{\rm h}^2 = \ap \lambda (\ha-\lambda)$.  This maps
into a configuration with seven D-branes at each fixed point and the other
two placed symmetrically.  The dilaton behaves as shown.

\vskip 1.5cm
\hskip 3.5cm\epsfxsize2.5in\epsfbox{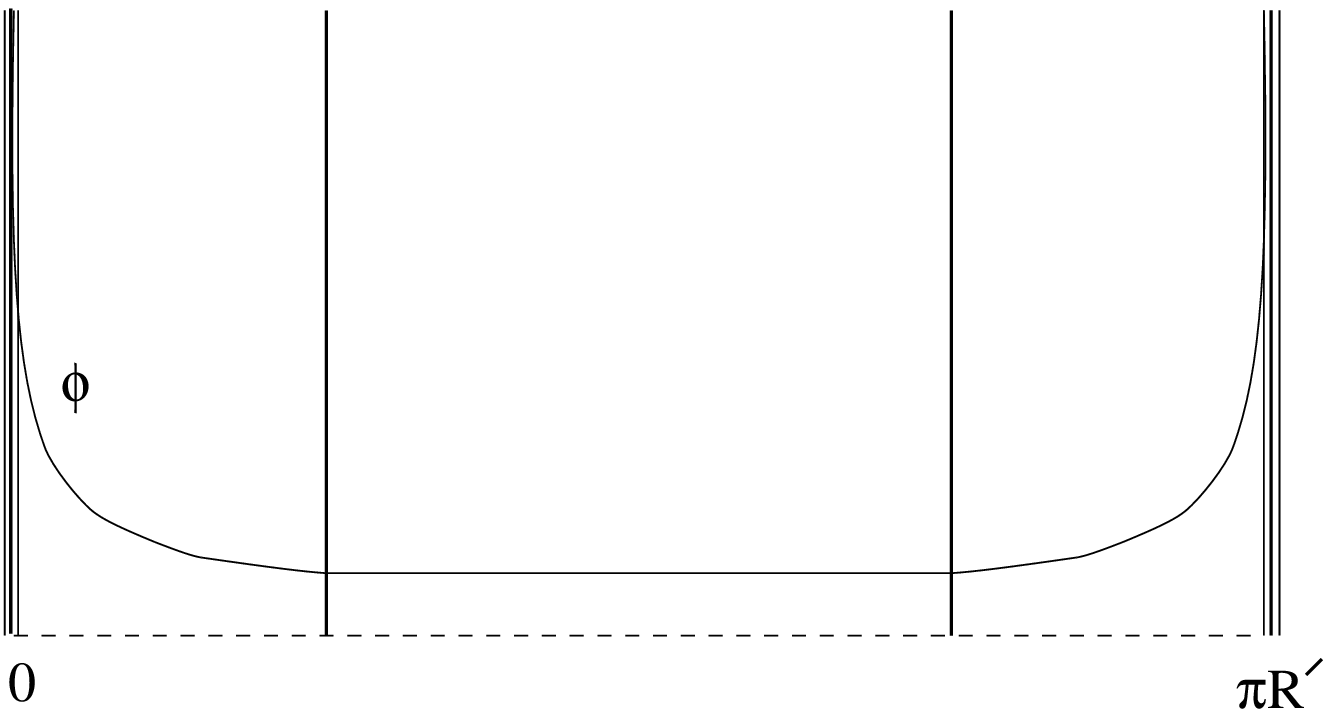}
\vskip 0.5cm

In the strong-coupling limit, the eighth D-brane moves toward each end.
In between, we have the IIa theory, in which an eleventh dimension is
supposed to decompactify at strong coupling.   The strong coupling limit is
then M-theory on $S_1/Z_2 \times S_1$, which is the same as  Horava and
Witten's description of the $E(8) \times E(8)$ M-theory~\ref\horwit{P.
Horava and E. Witten, {\it Heterotic and Type I String Dynamics from Eleven
Dimensions,} preprint IASSNS-HEP-95-86, hep-th/9510209.}.  Letting the
$S_1/Z_2$ have size $r_1$ and the $S_1$ size $r_2$, the dualities
give~\refs{\hullt,\wit,\horwit}
\eqn\mhi{\eqalign{
&R_{\rm h} = r_2 r_1^{1/2}, \qquad g_{\rm h} = r_1^{3/2} \cr
&R_{\rm I'} = r_1 r_2^{1/2}, \qquad g_{\rm I'} = r_2^{3/2} .}}
These agree with the mapping~\onephet\ between the Type I$'$ and the
$T$-dual of the heterotic theory.

\subsec{Five-Branes}

For the Type IIb five-brane we obtain again a world-brane $U(1)$ gauge field
plus the scalar transverse fluctuations, and their superpartners.  In the
Type I theory there is an interesting subtlety.  Consider multiple
five-branes, so the Chan-Paton index $i$ runs over both nine- and
five-branes.  The calculation~\cpcondition\ appears to imply that a common
$SO$ or $Sp$ projection must be taken on both types; we know from the
nine-brane that this must be $SO$.  However, in eq.~\cpcondition\ it was
assumed that $\Omega^2$ on the fields was simply the identity.  A more
careful analysis, explained in detail in section~2 of ref.~\ref\gimon{E.
Gimon and J. Polchinski, {\it Consistency Conditions for Orientifolds and
D-Manifolds,} preprint NSF-ITP-96-01, hep-th/9601038.}, shows that
$\Omega^2$ is $-1$ in the 5-9 sector of the open string Hilbert space.
To cancel this we must take the opposite projections on the five- and
nine-branes, so the former have a symplectic world-sheet group. In particular
this implies that they must appear in even numbers.\foot{This argument also
implies that Dirichlet three- and seven-branes are inconsistent in Type I
theory, as expected from the absence of an appropriate R-R field.}

Let us
therefore study two coincident five-branes.  The massless states are
\eqn\fives{
\lambda_{ij}\alpha^\mu_{-1}|{ k},ij\rangle, \qquad
\lambda'_{ij}\alpha^m_{-1}|{ k},ij\rangle
}
with $i,j$ running over $1,2$.
The orientation projection then implies that
\eqn\fivepro{
\epsilon \lambda = - \lambda^T \epsilon, \qquad
\epsilon \lambda' = \lambda^{\prime T} \epsilon.
}
This implies that $\lambda$ is one of the Pauli matrices, giving
five-brane gauge group $SU(2) = USp(2) = Sp(1)$.  The collective coordinate
wavefunction $\lambda'$ is the identity, so the two five-branes move as a
unit; physically it is a single five-brane with a two-valued Chan-Paton
factor.  This can also be seen in another way~\ref\witpriv{E. Witten, private
communication.}.  In the Type~I theory the force between 5-branes, and
between 1-branes, is half of what we found earlier, because of the
orientation projection on the sum over states. The product
of the charges of a single one-brane and single five-brane would then be
only half a Dirac-Teitelboim-Nepomechie unit; but since the five-branes
are always paired the quantization condition is respected.

This result, a symplectic gauge group on the five-brane, is required by
string duality~\ref\witsmall{E. Witten, {\it Small Instantons in String
Theory,} preprint IASSNS-HEP-95-87, hep-th/9511030.}.  The Type~I five-brane
is dual to the instanton five-brane of the heterotic theory.  The symplectic
gauge group is needed to give the correct moduli space of instantons.

In the Type~I theory there will also be 5-9 strings transforming as
a $({\bf 2},{\bf 32})$ under the five-brane and nine-brane gauge groups.
The R sector and NS sectors both have vanishing zero point energies, and in
each there are four periodic transverse fermions.  The zero modes thus
generate four states, reduced to two by the GSO projection.  In terms of
the $D=6$ $N=1$ supersymmetry of this configuration (equivalent to $D=4$,
$N=2$), this is the content of {\it half} a hypermultiplet.  This is
allowed because the representation~$({\bf 2},{\bf 32})$ is pseudoreal.

\subsec{A Brief Survey}

Dirichlet boundary conditions were a subject of frequent fascination
even before the relevance to string duality was realized, and were
interpreted in several different ways.  In this final section we will
briefly survey the pre- and post-duality literature on D-branes, omitting
some papers discussed elsewhere in these notes.

Boundaries with Dirichlet conditions on all coordinates (D-instantons or D
$(-1)$-branes in the current terminology) were first considered as
off-shell probes of the theory~\ref\offshell{J. H. Schwarz, Nucl. Phys.
{\bf B65} (1973), 131;\hfil\break E. F. Corrigan and D. B. Fairlie, Nucl.
Phys. {\bf B91} (1975) 527;\hfil\break M. B. Green, Nucl. Phys. {\bf B103}
(1976) 333;\hfil\break A. Cohen, G. Moore, P. Nelson, and J. Polchinski,
Nucl. Phys. {\bf B267}, 143 (1986); {\bf B281}, 127 (1987).}.
It was then proposed that introducing a gas of such boundaries would
produce the partonic behavior needed in a string theory of
QCD~\ref\parton{M. B. Green, Phys. Lett. {\bf B69} (1977) 89; {\bf
B201} (1988) 42; {\bf B282} (1992) 380; {\bf B329} (1994) 435.}.
In ref.~\joeone\ it was proposed that these D-instantons were actually an
essential part of  string theory, based on the $e^{-1/g}$ behavior noted in
ref.~\shenkscale.\foot
{Dirichlet boundary conditions are not superconformally invariant in
heterotic string theory, a fact which caused some discomfort in
ref.~\joeone.  D-branes have no known analog in the heterotic string, there
being no analog of the R-R fields, and so no understanding of the
$e^{-1/g}$ there.}

Boundaries with Dirichlet conditions on some coordinates and Neumann on
others were first suggested to represent a form of
compactification \ref\siegel{W. Siegel, Nucl. Phys. {\bf B109} (1976) 244.},
since the open strings move in a space of reduced dimension; however, the
closed strings still move in the critical dimension.  Such boundaries also
arose in an attempt to put the Type I string on a K3
orbifold~\ref\harvmin{J. A. Harvey and J. A. Minahan, Phys. Lett. {\bf 188B}
(1987) 44.}.  The interpretation in terms of a dynamical object was made in
ref.~\dbranes.
\nref\comps{N. Ishibashi and T. Onogi, Nucl. Phys. {\bf B318}
(1989) 239;\hfil\break Z. Bern, and D. C. Dunbar, Phys. Lett. {\bf B242}
(1990) 175; Phys. Rev. Lett. {\bf 64} (1990) 827;
Nucl. Phys. {\bf B319} (1989) 104; Phys. Lett. {\bf 203B} (1988)
109;\hfil\break M. Bianchi and A. Sagnotti, Phys. Lett. {\bf B247} (1990)
517; Nucl. Phys. {\bf B361} (1991) 519;\hfil\break A. Sagnotti, Phys. Lett.
{\bf B294} (1992) 196;
{\it Some Properties of Open-String
Theories,} preprint ROM2F-95/18, hep-th/9509080.}\nref\dabpark{
A. Dabholkar and J. Park, {\it An Orientifold of Type IIB Theory on K3,}
preprint CALT-68-2038, hep-th/9602030.}There were
many studies of compactification of open superstrings with orientifold
projections and Dirichlet boundaries, with spacetime anomaly and/or
divergence cancellation imposed~\refs{\harvmin,\ori,\comps}.  Recent
systematic studies of two examples can be found in
refs.~\refs{\gimon,\dabpark}.
\nref\vard{C. Vafa and E. Witten, {\it Dual String Pairs with $N=1$
and $N=2$ Supersymmetry in Four Dimensions,} HUTP-95-A023,
hep-th/9507050 (1995); \hfil\break
C. G. Callan, Jr. and I. R. Klebanov, {\it D-Brane Boundary State
Dynamics,} preprint PUPT-1578, hep-th/9511173;\hfil\break
S.-T. Yau and E. Zaslow, {\it BPS States, String Duality, and
Nodal Curves on K3,} preprint hep-th/9512121; \hfil\break R. G. Leigh,
{\it Anomalies, D-Flatness, and Small Instantons,} preprint RU-95-94,
hep-th/9512191.}

Many other interesting duality
properties of D-branes and orientifolds have recently been
discussed~\refs{\witbound,\bound,\csapp,\vard}.
There is no perturbative string theory in eleven dimensions, so our
knowledge of M theory is limited for now to some understanding of its
compactifications and duality symmetries.  The D-brane description of
R-R charges has been useful in unraveling some of this~\ref\mth{J.
Maharana, {\it M Theory $p$-Branes,} preprint IASSNS-HEP-95-98,
hep-th/9511159;\hfil\break
P. K. Townsend, {\it D-Branes from M-Branes,} preprint DAMTP-R-95-59,
hep-th/9512062;
A. Sen, {\it $T$-Duality of $p$-Branes,} preprint MRI-PHY-28-95,
hep-th/9512203;\hfil\break
E. Witten, {\it Five-Branes and M Theory on an Orbifold,}
preprint IASSNS-HEP-96-01, hep-th/9512219;\hfil\break
C. Schmidhuber, {\it D-Brane Actions,} preprint PUPT-1585,
hep-th/9601003;\hfil\break
J. H. Schwarz, {\it M Theory Extensions of
$T$-Duality,} preprint CALT-68-2034, hep-th/9601077.}.
\nref\scat{
I. R. Klebanov and Larus Thorlacius, {\it The Size of $p$-Branes,} preprint
PUPT-1574, hep-th/9510200;\hfil\break
S. S. Gubser, A. Hashimoto, I. R. Klebanov, and J.M. Maldacena,
{\it Gravitational Lensing by $p$-Branes,} preprint
PUPT-1586, hep-th/9601057;\hfil\break
J. L. F. Barbon, {\it D-Brane Form-Factors at High Energy,} preprint
PUPT-1590, hep-th/9601098.}

As mentioned above, the original interest in D-instantons was their hard
behavior at short distance.  Now that we are interpreting these as an
essential part of string theory we have to rethink this~\joeone: are D-branes
a sign of degrees of freedom at distances less than the string
scale~\shenkscale?  There have been several studies of
D-brane--D-brane and string--D-brane scattering.  The picture that
emerges is not entirely clear.  String scattering from a $p$-brane for $p
\geq 0$ has structure on the string scale~\scat, unlike D-instanton
corrections to scattering.  D-brane--D-brane scattering shows some sign of
shorter distance structure~\bachas.  The string scale structure has been
interpreted as a `string halo' that hides the shorter distance
physics~\shenkscale.  In fact, there is one other kind of scattering that
seems to cut through the string halo and see pointlike structure.  Consider
a macroscopic string ending on a D-brane.  We can send ripples down the
string and watch them bounce off the end.  This is easy: to lowest
order in string perturbation theory, the Dirichlet boundary condition just
gives an energy-independent phase shift, indicating a pointlike structure.
This holds up to arbitrarily high energies, so is cut off only where string
perturbation theory breaks down.  It will be interesting to pursue this
further.

\subsec{Conclusion}

So where does this leave us?  The goal is to answer the question ``What is
string theory?''  We have learned that string theory contains a new kind of
object, the D-brane, which is a sort of topological defect where strings
can end.  This has clarified many of the connections between dual theories,
and turned string duality into a much tighter structure.  However, it also
points up, even more strongly, that our current understanding of string
theory is only effective, provisional.  It is hard to imagine that string
theory will be defined in a precise way as some sort of sum over string {\it
and} D-brane world-sheets.  Rather, the perturbative string description is
valid only up to some scale, and the sum over D-brane histories makes no
sense at shorter distances.

We are in a position similar to that of Wilson~\ref\ken{
K. Wilson, Rev. Mod. Phys. {\bf 55} (1983) 583.}, when he was trying
to answer the question ``What is field theory?''
He began to make
progress when he found a model, the pion-nucleon static model, which was
simple enough to be understood yet rich enough to display the essence of
field theory.  We have finally found models, namely string backgrounds with
extended supersymmetry, which are simple enough that we can make progress,
but rich enough to display a great deal of new and surprising dynamics.
But the model was only a stepping-stone to the principle, which was to
think about field theory scale-by-scale.  That principle made possible both
a precise definition of field theory and an understanding of the dynamics
and phase structure.  In string theory we are still looking for the
underlying principle, and there is good reason to expect that it will be
similarly beautiful and powerful.

\listrefs
\bye